\documentclass[pra,twocolumn,tightenlines,showpacs,nofootinbib]{revtex4}

\usepackage{graphicx} 
\usepackage{amsmath}
\usepackage{bm}
\usepackage{dcolumn}

\begin{document}

\title{Relativistic all-order calculations of Th, Th$^{+}$ and Th$^{2+}$ atomic properties}
\author{ M. S. Safronova$^{1,2}$, U. I. Safronova$^3$, and Charles W. Clark$^2$}
\affiliation {$^1$Department of Physics and Astronomy, 217 Sharp
Lab, University of Delaware, Newark, Delaware 19716, \\
$^2$Joint Quantum Institute, National Institute of Standards and Technology
and the University of Maryland, Gaithersburg, Maryland 20899
\\$^3$Physics
Department, University of Nevada, Reno, Nevada 89557.
}

\date{\today}
\begin{abstract}
Excitation energies, term designations, and $g$-factors of
Th,  Th$^{+}$ and
Th$^{2+}$ are determined using a relativistic hybrid configuration interaction (CI) + all-order approach that combines configuration interaction
and linearized coupled-cluster methods. The results are compared with other theory and experiment where available. We find some ``vanishing'' $g$-factors, similar to those known in lanthanide spectra.  Reduced matrix elements, oscillator
strengths,  transition rates, and lifetimes are determined for  Th$^{2+}$.
To estimate  the uncertainties  of our results, we compared our values with the available
experimental lifetimes  for higher
$5f7p\ ^3G_{4}$, $7s7p\ ^3P_{0}$,
 $7s7p\ ^3P_{1}$, and $6d7p\ ^3F_{4}$ levels of Th$^{2+}$.
 These calculations provide
a benchmark test of the CI+all-order method for heavy systems with several valence electrons and yield
 recommended values for transition rates and
lifetimes of Th$^{2+}$.

\end{abstract}
\maketitle

\section{Introduction}

The $^{229}$Th  nucleus provides a unique opportunity  for the development of a nuclear clock ~\cite{peik2003} due to
an unusually low
first excitation energy of only several eV ~\cite{kroger,beck}, making the corresponding nuclear transition accessible with laser excitation \cite{ZhaEscRun12}.
 This very narrow, $6h$ lifetime \cite{ZhaEscRun12},  nuclear transition is expected to be well-isolated from effects of external fields leading to potentially very small ultimate uncertainty  in the corresponding frequency standard.
  The transition frequency is expected to be very sensitive to temporal variation of the fine-structure constant and the dimensionless strong interaction parameter $m_q/\Lambda_{\textrm{QCD}}$ as compared to atomic transitions, making $^{229}$Th one of the most attractive candidates for such studies \cite{flambaum06}.
The physical implementation of the frequency standard may employ, for example, the closed electronic shell of Th$^{4+}$ in a UV-transparent crystal doped with a macroscopic number of $^{229}$Th nuclei \cite{rellergert} or the stretched states within the $5f_{5/2}$ electronic ground level of both nuclear ground and isomer manifolds of a single trapped ion ~\cite{CamRadKuz12}.
 Laser-cooled Wigner crystals of $^{229}$Th$^{3+}$ allow for high-precision spectroscopy  \cite{gt3,RadCamKuz12}. Singly- and doubly-charged $^{232}$Th and $^{229}$Th ions have been
produced by laser ablation of solid-state thorium compounds and by
inductively coupled plasma techniques with mass-spectrometry
analysis from liquid solutions of thorium \cite{TroBorKha13}. The
latter method was found to be more applicable for producing ions
of radioactive $^{229}$Th for laser experiments when searching for
the energy value of the isomeric nuclear transition \cite{TroBorKha13}.

We note that a frequency standard based on the nuclear transition in $^{229}$Th can be implemented with either neutral or ionized Th. The nuclear transition might be accessed using an electronic bridge process \cite{PorFlaPei10}, which involves matching
combined electronic and nuclear energy levels to drive the nuclear transition \cite{PorFlaPei10}. Implementing the
electronic bridge will  require knowledge of the complex electronic configurations of neutral or ionized Th.

Much of the recent work on the spectrum of thorium is motivated by its use
as source of wavelength standards for high-resolution spectrographs ~\cite{RedNavSan14}.
  The high density of Th~I spectral lines - approximately 20~000 in the wavelength
  interval between 250~nm and 5500~nm - has made the thorium-argon hollow-cathode lamp
  a convenient tool for accurate wavelength calibration. In particular,
  such lamps  are installed  on many high-precision astronomical spectrographs. These include
  the High Accuracy Radial velocity Planet Searcher (HARPS) instrument, now
the world's most precise astronomical
spectrograph, which has a relative precision of 3 parts in $10^9$, and can measure stellar velocities  to within  $\sim$1~m/s \cite{HARPS}.

The best characterized Th ion today is  Th$^{3+}$, which has monovalent Fr-like electronic structure and has been studied with the high-precision all-order method ~\cite{SafSaf13}. Recommended values for  electric-dipole matrix elements, oscillator strengths,
transition rates, lifetimes,  scalar and tensor polarizabilities, and hyperfine constants critically
evaluated   for their accuracy have been published  for a large number states in Th$^{3+}$ ~\cite{SafSaf13}. The combination of the experimental measurements of hyperfine constants of Th$^{3+}$ with theoretical calculations has enabled accurate determination of
magnetic dipole and electric quadrupole moments
\cite{SafSafRad13}.
Analysis of resonant excitation
Stark ionization spectroscopy spectra has led to the determination of
 Th$^{3+}$ ground-state electric quadrupole moment,
adiabatic scalar and tensor dipole polarizabilities, and the dipole
matrix elements connecting the ground  level to
low-lying excited levels of Th$^{3+}$ \cite{lundeen}.

Due to the more complicated atomic structure of Th$^{2+}$, Th$^{+}$ and Th, which have respectively two, three and four valence electrons,  their atomic properties are less precisely known than that of Th$^{3+}$. The work presented in this paper demonstrates accurate calculation  of the energies  of these systems and provides
transition matrix elements, oscillator strengths, and lifetimes for a large number of Th$^{2+}$ states. This work also
serves as  a benchmark test of the accuracy of the CI+all-order method for systems with multiple valence electrons, including tetravalent neutral Th.

\section{Review of current knowledge of structure of thorium and its ions}
Properties of thorium and its ions have received more
extensive experimental than theoretical investigation, due to significant difficulties in accurate first-principles calculations of such heavy many-electron systems.
We begin with a summary of the experimental work and conclude with a survey
 of theoretical approaches.

A list of about 9500 spectral lines of Th in the
range 234.5-2966.2~nm  ~was obtained and characterized by Zalubas
\cite{Zalubas76}. This resulted in determination of  254 even-parity and 322 odd-parity levels. Their $g$-factors were  obtained by Zeeman spectroscopy, and used as an aid
    in spectral classification, i.e. the assignment of angular momenta, parity,  and electronic configurations.

 Energy levels and classified lines in the second
spectrum of thorium (Th II) were described by Zalubas and Corliss
\cite{ZalCor74}. About 6500 lines were classified as transitions
between 199 odd levels and 271 even levels; 188 levels
result from the  odd $5f6d^{2} + 5f7s^2$ and
($6d^{2}7p + 7s^{2}7p$) configurations \cite{ZalCor74}.
  The 235 levels  of the even
$5f^27s + 5f7s7p + 5f6d7p + 5f^26d$ configurations were also determined
 \cite{ZalCor74}.  Resonantly enhanced
three-photon ionization of Th$^+$ was used to determine that its ionization potential
is between 11.9~eV
and 12.3~eV \cite{HerNemOkh13}.

 The Th~III spectrum was observed in the region 1000-3000~nm, and ten lines were identified \cite{Lit74}. The ground level of Th$^{2+}$ was determined to be
 $5f6d\ ^3H_4$ \cite{Lit74}. The first excited level, $6d^2\ ^3F_2$,
was determined to be only 63.2~cm$^{-1}$ above the ground level. Wyart and
Kaufman \cite{WyaKau81} extended the
analysis of Th$^{2+}$. They classified the 92  lines above
194~nm that were previously observed~\cite{WyaKau81}.  The
$5f6d$, $5f^2$, $5f7d$, and $5f8s$ configurations were completely identified
\cite{WyaKau81}.

 Using time-resolved laser-induced
fluorescence method, Bi\'emont {\it et al.\/} measured the lifetimes of six levels belonging to the
$5f^2$, $5f7p$, $7s7p$, and $6d7p$ configurations of Th$^{2+}$\cite{BiePalQui02}.
 These transitions provide the mechanism of a
cosmochronometer  for estimating
the age of the Galaxy.

An online
database of published and unpublished actinide energy levels
\cite{Thorium} lists 693 levels of Th and 507 of Th$^+$ 
with uncertainties of 0.001~cm$^{-1}$.
 Recently, Redman {\it et al.\/} \cite{RedNavSan14}
presented results of precise observations of a thorium–-argon
hollow cathode lamp emission spectrum in the region between 3500~nm
and 1175~nm using a high-resolution Fourier transform
spectrometer. Their  measurements were combined with results from
 previously published thorium line lists \cite{Sansonetti-84,Sansonetti-02,Sansonetti-05,Sansonetti-06,Sansonetti-08,Sansonetti-12}
 to re-optimize the
energy levels of neutral, singly, and doubly ionized thorium
(Th,  Th$^{+}$, and Th$^{2+}$).
 A systematic analysis of previous
measurements in light of  these new results enabled [Redman {\it et al.\/}
 \cite{RedNavSan14}] to
identify and propose corrections for systematic errors and
typographical errors and incorrect classifications in previous
identifications. Redman {\it et al.\/} \cite{RedNavSan14} present 787 levels of Th I and 516 of Th II in their tables.

Dzuba and Flambaum \cite{DzuFla10} presented analytical estimates
and numerical calculations showing that the energy level density
in open-shell atoms increases exponentially with
excitation energy. They used the relativistic
Hartree-Fock and configuration interaction methods to calculate
the densities of states of Th and Th$^{+}$. Their results were used to
estimate the effect of electrons on the nuclear clock transition  discussed
in the previous section \cite{DzuFla10}.

Porsev and Flambaum \cite{PorFla10} used the CI+many-body perturbation theory (MBPT) method to study the effect of atomic
electrons on the nuclear clock transition  due to the
electronic bridge process. They   calculated energies of
several high-lying even-parity states that have yet to seen by experiments.

Roy {\it et al.\/} have performed relativistic two-component {\it ab initio} calculations
for  Th$^{+}$ and Th$^{2+}$ ions
\cite{RoyPraDat12}.

In our present work, we evaluate atomic properties of Th,
Th$^{+}$, and Th$^{2+}$ using the CI+all-order approach.
Excitation energies and $g$-factors are compared with  experimental \cite{Thorium}
and other theoretical results \cite{BerDzuFla09}. We also
evaluate multipole transition rates and lifetimes of low-lying
levels for Th$^{2+}$.

 \begin{table*}
\caption{\label{th1}
 Levels (cm$^{-1}$) and
$g$-factors of the lowest states of
  thorium. Non-relativistic values of $g$-factors ($g_{\rm nr}$)
  are given by Eq.~(\ref{eq-gnr}).
    Comparison of calculations with experiment \cite{Thorium}. Configuration and term labels are determined as described in Section~\ref{term}.}
\begin{ruledtabular}
\begin{tabular}{llllrrrrrllllrrrrrr}
\multicolumn{1}{l}{Conf.}&
\multicolumn{3}{c}{Term}&
\multicolumn{2}{c}{Energy}&
\multicolumn{3}{c}{g-factor}&
\multicolumn{1}{c}{Conf.}&
\multicolumn{3}{c}{Term}&
\multicolumn{2}{c}{Energy}&
\multicolumn{3}{c}{g-factor}\\
\multicolumn{1}{c}{}&
\multicolumn{1}{c}{\cite{Thorium}}&
\multicolumn{1}{c}{Present}&
\multicolumn{1}{c}{$J$}&
\multicolumn{1}{c}{Present}&
\multicolumn{1}{c}{Expt. }&
\multicolumn{1}{c}{Expt.}&
\multicolumn{1}{c}{Present}&
\multicolumn{1}{c}{nr}&
\multicolumn{1}{c}{}&
\multicolumn{1}{c}{\cite{Thorium}}&
\multicolumn{1}{c}{Present}&
\multicolumn{1}{c}{$J$}&
\multicolumn{1}{c}{Present}&
\multicolumn{1}{c}{Expt.}&
\multicolumn{1}{c}{Expt.}&
\multicolumn{1}{c}{Present}&
\multicolumn{1}{c}{nr}\\
\hline
\multicolumn{9}{c}{Even-parity states ($6d^27s^{2}+6d^37s$)}&
\multicolumn{9}{c}{Odd-parity states ($6d^27s7p+ 5f6d^27s + 5f6d7s^{2} +6d7s^27p $)}\\
\hline
   $  6d^27s^2  $&$ ^3F$ &$  ^3F$ &2&       0&      0&  0.735 &   0.718 &0.667   &   $  5f6d7s^2  $&$ ^3P$ &$     $ & 0&  13954& 14247 & 0.0    &   0.000 &   0.0  \\
   $  6d^27s^2  $&$    $ &$  ^3P$ &2&   3790 &  3688 & 1.255  &   1.287 & 1.500  &   $  6d^27s7p  $&$ ^5D$ &$     $ & 0&  18127& 18382 & 0.0    &   0.000 &   0.0  \\
   $  6d^37s    $&$ ^5F$ &$  ^5F$ &2&   6677 &  6362 & 1.010  &   1.001 & 1.000  &   $  6d7s^27p  $&$ ^3P$ &$     $ & 0&  20790& 20543 & 0.0    &   0.000 &   0.0  \\
   $  6d^27s^2  $&$    $ &$  ^3D$ &2&   7445 &  7280 & 1.185  &   1.152 & 1.167  &                 &       &        &  &       &       &        &         &        \\
   $  6d^37s    $&$ ^5P$ &$  ^5P$ &2&  12449 & 11802 & 1.780  &   1.759 & 1.833  &   $  6d7s^27p  $&$ ^3D$ &$  ^3D$ & 1&  11455& 11878 & 0.725  &   0.709 &  0.500 \\
                &        &         &&        &       &        &         &        &   $  5f6d7s^2  $&$ ^3P$ &$  ^3P$ & 1&  13989& 14244 & 1.205  &   1.218 &  1.500 \\
   $  6d^27s^2  $&$ ^3P$ &$     $ &0 &   2708&  2558 & 0.0    &   0.000 &  0.0   &   $  6d^27s7p  $&$ ^5F$ &$  ^5F$ & 1&  15805& 15737 & 0.385  &   0.297 &  0.000 \\
   $  6d^37s    $&$    $ &$     $ &0 &  15243& 14227 & 0.0    &   0.000 &  0.0   &   $  5f6d7s^2  $&$ ^3D$ &$  ^1P$ & 1&  17388& 17357 & 0.505  &   1.024 &  1.000 \\
   $  6d^27s^2  $&$    $ &$     $ &0 &  17327& 16351 & 0.0    &   0.000 &  0.0   &                 &       &        &  &       &       &        &         &        \\
                 &        &         &&        &      &        &         &        &   $  5f6d7s^2  $&$ ^3F$ &$  ^3F$ & 2&   7493&  8244 & 0.775  &   0.796 &  0.667 \\
   $  6d^27s^2  $&$ ^3P$ &$  ^3P$ &1&    3948&  3865 &1.480   &   1.481 &  1.500 &   $  6d7s^27p  $&$ ^3F$ &$  ^3F$ & 2&  10481& 10783 & 0.725  &   0.730 &  0.667 \\
   $  6d^37s    $&$ ^5F$ &$  ^5F$ &1&    5887&  5563 &0.065   &   0.048 &  0.000 &   $  5f6d7s^2  $&$ ^1D$ &$  ^1D$ & 2&  11223& 12114 & 0.975  &   0.921 &  1.000 \\
   $  6d^37s    $&$ ^5P$ &$  ^5P$ &1&   12179& 11601 &2.400   &   2.430 &  2.500 &   $  6d7s^27p  $&$ ^3D$ &$  ^3D$ & 2&  13641& 14032 & 1.125  &   1.239 &  1.167 \\
   $  6d^37s    $&$    $ &$  ^1P$ &1&   14754& 13963 &0.760   &   0.714 &  1.000 &   $  6d^27s7p  $&$ ^5G$ &$  ^3F$ & 2&  14253& 14465 & 0.810  &   0.759 &  0.667 \\
                  &        &       &&        &       &        &         &        &   $  6d7s^27p  $&$ ^1D$ &$  ^1D$ & 2&  15879& 16217 & 1.070  &   1.085 &  1.000 \\
   $  6d^27s^2  $&$ ^3F$ &$  ^3F$ &3&    2815&  2869 &1.085   &   1.078 &  1.083 &   $  6d^27s7p  $&$ ^5F$ &$  ^5F$ & 2&  17187& 17224 & 1.045  &   1.029 &  1.000 \\
   $  6d^37s    $&$ ^5F$ &$  ^5F$ &3&    7818&  7502 &1.250   &   1.241 &  1.250 &   $  5f6d7s^2  $&$ ^3P$ &$  ^3D$ & 2&  17781& 17847 & 1.165  &   1.122 &  1.167 \\
   $  6d^37s    $&$ ^5P$ &$  ^5P$ &3&   13413& 12848 &1.390   &   1.629 &  1.667 &                 &       &        &  &       &       &        &         &        \\
   $  6d^37s    $&$ ^3G$ &$  ^3G$ &3&   13613& 13089 &1.050   &   0.800 &  0.750 &   $  5f6d7s^2  $&$ ^3G$ &$  ^5G$ & 3&  10194& 10527 & 0.870  &   0.940 &  0.917 \\
   $  6d^37s    $&$    $ &$  ^3F$ &3&   16685& 15970 &1.205   &   1.166 &  1.083 &   $  5f6d7s^2  $&$ ^3F$ &$  ^1F$ & 3&  10762& 11242 & 1.010  &   0.914 &  1.000 \\
   $  6d^37s    $&$    $ &$  ^5F$ &3&   18138& 17398 &1.195   &   1.213 &  1.250 &   $  6d7s^27p  $&$ ^3F$ &$  ^3F$ & 3&  13664& 13945 & 1.110  &   1.129 &  1.083 \\
   $  6d^37s    $&$ ^3F$ &$  ^3F$ &3&   20527& 19713 &1.110   &   1.094 &  1.083 &   $  6d^27s7p  $&$ ^5G$ &$  ^3G$ & 3&  15054& 15167 & 1.065  &   0.591 &  0.750 \\
   $  6d^37s    $&$ ^1F$ &$  ^5F$ &3&   22222& 21595 &1.040   &   1.211 &  1.250 &                 &       &        &  &       &       &        &         &        \\
                 &       &         &&        &       &        &         &        &   $  5f6d7s^2  $&$ ^3H$ &$  ^3H$ & 4&   7296&  7795 & 0.865  &   0.872 &  0.800 \\
   $  6d^27s^2  $&$ ^3F$ &$  ^3F$ &4&    4953&  4962 &1.210   &   1.219 &  1.250 &   $  5f6d7s^2  $&$ ^1G$ &$  ^1G$ & 4&   9858& 10414 & 0.985  &   0.962 &  1.000 \\
   $  6d^27s^2  $&$ ^1G$ &$  ^5G$ &4&    8156&  8111 &1.065   &   1.213 &  1.150 &   $  5f6d7s^2  $&$ ^3G$ &$  ^3G$ & 4&  12827& 13175 & 1.095  &   1.111 &  1.050 \\
   $  6d^37s    $&$ ^5F$ &$  ^5G$ &4&    9113&  8800 &1.310   &   1.138 &  1.150 &   $  5f6d7s^2  $&$ ^3F$ &$  ^3F$ & 4&  13683& 14207 & 1.170  &   1.135 &  1.250 \\
   $  6d^37s    $&$ ^3G$ &$  ^1G$ &4&   13782& 13297 &1.000   &   0.990 &  1.000 &   $  5f6d^27s  $&$ ^5H$ &$  ^5H$ & 4&  16092& 16347 & 0.880  &   0.939 &  0.900 \\
   $  6d^37s    $&$ ^3H$ &$  ^3H$ &4&   15939& 15493 &0.905   &   0.887 &  0.800 &   $  5f6d^27s  $&$ ^5I$ &$  ^5I$ & 4&  16840& 16784 & 0.695  &   0.639 &  0.600 \\
   $  6d^37s    $&$ ^3F$ &$  ^3F$ &4&   18692& 17960 &1.175   &   1.209 &  1.250 &   $  6d7s^27p  $&$ ^3F$ &$  ^3F$ & 4&  17883& 18054 & 1.185  &   1.165 &  1.250 \\
   $  6d^37s    $&$ ^3F$ &$  ^3F$ &4&   20347& 19532 &1.204   &   1.196 &  1.250 &   $  6d^27s7p  $&$ ^5G$ &$  ^5G$ & 4&  18503& 18810 & 1.150  &   1.129 &  1.150 \\
   $  6d^37s    $&$    $ &$  ^3G$ &4&   21725& 21646 &1.090   &   1.056 &  1.050 &                 &       &        &  &       &       &        &         &        \\
                 &       &        &&         &       &        &         &        &   $  5f6d7s^2  $&$ ^3H$ &$  ^3H$ & 5& 10884 & 11197 & 1.040  &   1.028 &  1.033 \\
   $  6d^37s    $&$ ^5F$ &$  ^5F$ &5&   10198&  9805 &1.365   &   1.360 &  1.400 &   $  5f6d7s^2  $&$ ^3G$ &$  ^3G$ & 5& 15255 & 15490 & 1.190  &   1.188 &  1.200 \\
   $  6d^37s    $&$    $ &$  ^3G$ &5&   14723& 14204 &1.150   &   1.134 &  1.200 &   $  5f6d^27s  $&$ ^5I$ &$  ^5H$ & 5& 17311 & 17501 & 1.015  &   1.099 &  1.100 \\
   $  6d^37s    $&$ ^3H$ &$  ^3H$ &5&   17614& 17166 &1.115   &   1.111 &  1.033 &   $  5f6d^27s  $&$ ^5H$ &$  ^1H$ & 5& 17974 & 18011 & 1.025  &   0.942 &  1.000 \\
   $  6d^37s    $&$ ^1H$ &$  ^1H$ &5&   21775& 21143 &1.030   &   1.005 &  1.000 &   $  6d^27s7p  $&$ ^5G$ &$  ^5H$ & 5& 19320 & 19588 & 1.150  &   1.151 &  1.100 \\
   $ 5f6d7s7p   $&$ ^5I$ &$  ^1H$ &5&   22845& 23277 &1.010   &   1.007 &  1.000 &   $  5f6d^27s  $&$ ^1H$ &$  ^3H$ & 5& 20586 & 20322 & 1.060  &   1.040 &  1.033 \\
   $ 5f6d7s7p   $&$ ^5H$ &$  ^3H$ &5&   26060& 26381 &1.025   &   1.031 &  1.033 &   $  6d^27s7p  $&$ ^5F$ &$  ^5G$ & 5& 20932 & 21077 & 1.250  &   1.246 &  1.267 \\
                 &       &        &&         &       &        &         &        &   $  5f6d^27s  $&$ ^3I$ &$  ^5I$ & 5& 22537 & 22399 & 0.930  &   0.893 &  0.900 \\
   $  6d^37s    $&$  ^3H$ &$ ^3H$ &6&   17063& 16554 &1.165   &   1.161 &  1.167 &                 &       &        &  &       &       &        &         &        \\
   $  5f6d7s7p  $&$  ^5I$ &$ ^5I$ &6&   26681& 26997 &1.110   &   1.109 &  1.071 &   $  5f6d7s^2  $&$ ^3H$ &$  ^3H$ & 6&  14188& 14482 & 1.170  &   1.160 &  1.167\\
   $  6d^4      $&$  ^3H$ &$ ^3H$ &6&   28926& 27972 &1.125   &   1.164 &  1.167 &   $  5f6d^27s  $&$ ^5I$ &$  ^5I$ & 6&  19332& 19227 & 1.085  &   1.069 &  1.071\\
   $  5f6d7s7p  $&$  ^5H$ &$ ^3H$ &6&   29335& 29553 &1.185   &   1.139 &  1.167 &   $  5f6d^27s  $&$ ^5H$ &$  ^5H$ & 6&  19940& 19986 & 1.195  &   1.192 &  1.214\\
                 &        &        &&        &       &        &         &        &   $  5f6d^27s  $&$ ^3I$ &$  ^3I$ & 6&  23282& 23307 &        &   1.029 &  1.024\\
                 &        &        &&        &       &        &         &        &   $  6d^27s7p  $&$ ^5G$ &$  ^5G$ & 6&  24038& 24085 &1.220   &   1.260 &  1.333\\
                 &        &        &&        &       &        &         &        &   $  5f6d7s^2  $&$ ^3I$ &$  ^3H$ & 6&  24861& 24850 &        &   1.179 &  1.167\\
 \end{tabular}
\end{ruledtabular}
\end{table*}

\section{Computational Method}

Calculation of the properties of  thorium and it first few ions requires accurate all-order treatment of electron correlations. Low-order perturbation methods are ineffective
in heavy systems with more than one valence electron, due to large effects
of  valence-valence electronic correlations. Moreover, the radon-like core of the thorium atom is sufficiently large that core-core and core-valence correlations have to be treated accurately as well. This can be
accomplished within the framework of the CI+all-order method  that combines configuration interaction and
coupled-cluster approaches \cite{SafKozJoh09,SafKozCla11,PorSafKoz12,SafPorKoz12,SafPorCla12}.
The CI+all-order method was used to evaluate properties of
systems with three valence electrons
in Refs.~\cite{PorSafKoz12a,SafSafPor13,SafMaj13,SafDzuFla14}.
In 2014, properties of systems with  four valence electrons were
calculated with the CI+all-order method in Ref.~\cite{SafDzuFla14} for Sn-like ions.
The  spectra of the
superheavy elements No, Lr and Rf with two, three, and four valence electrons
was recently presented by Dzuba {\it et
al.\/} \cite{DzuSafSaf14}.

In the CI+all-order method, we start with a solution of the Dirac-Fock  (DF) equations
 \begin{equation}
  H_0\, \psi_c = \varepsilon_c \,\psi_c,
  \end{equation}
  where
$H_0$ is the relativistic DF Hamiltonian \cite{DzuFlaKoz96b,SafKozJoh09}
 and $\psi_c$ and
$\varepsilon_c$ are single-electron wave functions and energies. The calculations are
 carried out in the
$V^{\rm{N-4}}$, $V^{\rm{N-3}}$, and $V^{\rm{N-2}}$ potentials for Th, Th$^+$, and Th$^{2+}$, respectively, where $N$
is the total number of the electrons.
Therefore, the calculations are carried out with
the same radon-like Th$^{4+}$ frozen-core Dirac-Fock potential of all three systems considered in this work.

The wave functions and the corresponding low-lying energy
levels are determined by solving the many-electron relativistic
equation for two, three, or four valence electrons~\cite{KotTup87}, $$ H_{\rm
eff}(E_n) \Phi_n = E_n \Phi_n. $$ The effective Hamiltonian is
defined as
\begin{equation}
 H_{\rm eff}(E) = H_{\rm FC} + \Sigma(E),
 \end{equation}
  where
$H_{\rm FC}$ is the Hamiltonian in the frozen-core approximation.
 The energy-dependent effective Hamiltonian term $\Sigma(E)=\Sigma _1+\Sigma
_2$ is calculated using a modified version of the all-order linearized
coupled-cluster method with single and double excitations (LCCSD)
described in \cite{mar-pol-99,safr-ca-11}.
Therefore, the effective Hamiltonian contains dominant core and
core-valence correlation corrections to all orders.
 The valence correlations are treated by the CI method ~\cite{KotTup87}.
We refer the reader to Ref.~\cite{SafKozJoh09}
for the formulas and detailed description of the CI + all-order
method.

 The CI + all-order approach is based
on the Brillouin-Wigner variant of the many-body perturbation
theory, rather than the Rayleigh-Schr\"{o}dinger variant, leading to dependence of $\Sigma$ upon energy. This introduces some subtleties associated with appropriate
treatment of energy denominators. This issue has been discussed in Ref.~\cite{SafKozJoh09}
and we adopt the technical procedure that was recommended there.

The configurations are strongly mixed in all three systems considered here.
We present the results for the following configurations:
\begin{itemize}
\item Th: $6d^27s^{2}$, $6d^37s$, $6d^4$, $6d^27s7p$, $5f6d^27s$, $5f6d7s^{2}$,
 $6d7s^27p$;
\item Th$^{+}$: $6d7s^2$, $6d^27s$, $6d^{3}$, $5f^27s$, $5f^26d$, $5f6d7s$, $5f7s^{2}$, $5f6d^2$;
\item Th$^{2+}$: $5f6d$, $5f7d$, $5f7s$, $5f8s$, $6d6f$, $6d7p$,
$6d^{2}$, $5f^{2}$, $7s^{2}$, $5f7p$, $5f6f$, $6d7s$.
\end{itemize}
We also calculate $g$-factors and  compare them with
experimental values given in Ref.~\cite{Thorium}. For a
single configuration that is described by
pure $LS$-coupling, the non-relativistic
$g$-factor of the many-electron state is given by the Land\`e formula
\begin{equation}
  g_{\rm nr} = 1 + \frac{J(J+1)-L(L+1)+S(S+1)}{2J(J+1)},
\label{eq-gnr}
\end{equation}
where $J$ is the total angular momentum, $L$ is the total orbital angular
momentum and $S$ is the total spin angular momentum.
As will be discussed
  in the next section, we find that the non-relativistic
 $g$-factors are often useful for spectral term classification.


 \begin{table*}
\caption{\label{th2}
 Levels (cm$^{-1}$) and
$g$-factors of the lowest states of one-time ionized
  thorium. Non-relativistic values of $g$-factors ($g_{\rm nr}$)
  are given by Eq.(\ref{eq-gnr}).
    Comparison of calculations with experiment \cite{RedNavSan14}. Configuration and term labels are determined as described in Section~\ref{term}.}
\begin{ruledtabular}
\begin{tabular}{llllrrrrrllllrrrrrr}
\multicolumn{1}{c}{Conf.}&
\multicolumn{3}{c}{Term}&
\multicolumn{2}{c}{Energy}&
\multicolumn{3}{c}{g-factor}&
\multicolumn{1}{c}{Conf.}&
\multicolumn{3}{c}{Term}&
\multicolumn{2}{c}{Energy}&
\multicolumn{3}{c}{g-factor}\\
\multicolumn{1}{c}{}&
\multicolumn{1}{c}{\cite{Thorium}}&
\multicolumn{1}{c}{Present}&
\multicolumn{1}{c}{$J$}&
\multicolumn{1}{c}{Present}&
\multicolumn{1}{c}{Expt.}&
\multicolumn{1}{c}{Expt.}&
\multicolumn{1}{c}{Present}&
\multicolumn{1}{c}{nr}&
\multicolumn{1}{c}{}&
\multicolumn{1}{c}{\cite{Thorium}}&
\multicolumn{1}{c}{Present}&
\multicolumn{1}{c}{$J$}&
\multicolumn{1}{c}{Present}&
\multicolumn{1}{c}{Expt.}&
\multicolumn{1}{c}{Expt.}&
\multicolumn{1}{c}{Present}&
\multicolumn{1}{c}{nr}\\
\hline
\multicolumn{9}{c}{Even-parity states ($6d^27s+6d^{3}+5f^27s+5f^26d$)}&
\multicolumn{9}{c}{Odd-parity states ($5f6d7s + 5f7s^{2} + 5f6d^2 $)}\\
\hline
$ 6d^27s $&$     $&$  ^2D $&3/2 &     0 &      0  &  0.639  &   0.662 &    0.800&  $ 5f6d7s $&$ ^4D$&$  ^4D  $& 1/2 & 11550 &    11725  &  0.255  &  0.239 &  0.000 \\
$ 6d^27s $&$ ^4F $&$  ^4F $&3/2 &  1948 &   1859  &  0.586  &   0.554 &    0.400&  $ 5f6d7s $&$    $&$  ^2P  $& 1/2 & 13965 &    14102  &  0.523  &  0.552 &  0.667 \\
$ 6d^3   $&$     $&$  ^2D $&3/2 &  7779 &   7001  &  0.800  &   0.806 &    0.800&  $ 5f6d7s $&$ ^4P$&$  ^4P  $& 1/2 & 15401 &    15324  &  2.565  &  2.567 &  2.667 \\
$ 6d^27s $&$ ^4P $&$  ^4P $&3/2 &  8622 &   8018  &  1.608  &   1.608 &    1.733&  $ 5f6d7s $&$    $&$  ^2S  $& 1/2 & 17538 &    17838  &  1.08   &  1.582 &  2.000 \\
$ 6d^3   $&$     $&$  ^2D $&3/2 &  9209 &   8460  &  0.968  &   0.945 &    0.800&            &             &         &     &       &           &         &              \\
$ 6d^27s $&$ ^2D $&$  ^4D $&3/2 & 13270 &  12220  &  0.977  &   0.946 &    1.200&  $ 5f6d7s $&$ ^4F$&$  ^4F  $& 3/2 &  6020 &     6691  &  0.492  &  0.510 &  0.400 \\
$ 6d^3   $&$ ^4P $&$  ^4P $&3/2 & 16503 &  15237  &  1.592  &   1.612 &    1.733&  $ 5f6d7s $&$ ^2D$&$  ^2D  $& 3/2 & 10695 &    11576  &  0.832  &  0.754 &  0.800 \\
$ 6d^3   $&$     $&$  ^2D $&3/2 & 19973 &  18119  &  0.93   &   0.862 &    0.800&  $ 5f6d7s $&$    $&$  ^4D  $& 3/2 & 12657 &    12902  &  1.167  &  1.184 &  1.200 \\
          &       &        &    &       &         &         &         &         &  $ 5f6d7s $&$    $&$  ^2P  $& 3/2 & 15198 &    15145  &  1.366  &  1.278 &  1.333 \\
$ 6d^27s $&$ ^4P $&$  ^4P $&1/2 &  6972 &   6244  &  2.112  &   2.144 &    2.667&  $ 5f6d7s $&$ ^2D$&$  ^4D  $& 3/2 & 15676 &    15711  &  1.06   &  1.162 &  1.200 \\
$ 6d^27s $&$     $&$  ^2S $&1/2 &  8509 &   7828  &  1.254  &   1.201 &    2.000&            &             &         &     &       &           &         &              \\
$ 6d^3   $&$ ^4P $&$  ^4P $&1/2 & 15632 &  14349  &  2.555  &   2.564 &    2.667&  $ 5f7s^2 $&$ ^2F $&$  ^2F  $& 5/2 &  3882 &     4490  &  0.856  &  0.853 &  0.857 \\
          &       &        &    &       &         &         &         &         &  $ 5f6d7s $&$ ^4F $&$  ^4F  $& 5/2 &  6650 &     7331  &  1.061  &  1.070 &  1.029 \\
$ 6d^27s $&$ ^4F $&$  ^4F $&5/2 &  1722 &   1522  &  1.076  &   1.070 &    1.029&  $ 5f6d7s $&$ ^4G $&$  ^4G  $& 5/2 &  9299 &     9585  &  0.601  &  0.606 &  0.571 \\
$ 6d7s^2 $&$     $&$  ^2D $&5/2 &  4185 &   4113  &  1.163  &   1.150 &    1.200&  $ 5f6d7s $&$     $&$  ^4F  $& 5/2 &  9978 &    10673  &  1.088  &  1.089 &  1.029 \\
$ 6d^27s $&$ ^2F $&$  ^2F $&5/2 &  9198 &   8606  &  0.986  &   0.982 &    0.857&  $ 5f6d7s $&$ ^2F $&$  ^4F  $& 5/2 & 12045 &    12472  &  0.982  &  0.935 &  1.029 \\
$ 6d^27s $&$ ^4P $&$  ^4P $&5/2 &  9680 &   9061  &  1.419  &   1.408 &    1.600&  $ 5f6d7s $&$ ^4D $&$  ^4D  $& 5/2 & 14310 &    14546  &  1.339  &  1.346 &  1.371 \\
$ 6d^3   $&$     $&$  ^4F $&5/2 & 10145 &   9401  &  1.034  &   1.035 &    1.029&            &             &         &     &       &           &         &              \\
$ 6d^27s $&$     $&$  ^2D $&5/2 & 14220 &  13251  &  1.245  &   1.235 &    1.200&  $ 5f6d7s $&$ ^4H $&$  ^4H  $& 7/2 &  5743 &     6168  &  0.718  &  0.729 &  0.667 \\
$ 6d^3   $&$ ^4P $&$  ^4P $&5/2 & 17100 &  15787  &  1.571  &   1.566 &    1.600&  $ 5f7s^2 $&$ ^2F $&$  ^2F  $& 7/2 &  7794 &     8379  &  1.132  &  1.127 &  1.143 \\
$ 6d^3   $&$ ^2D $&$  ^4D $&5/2 & 22144 &  20159  &  1.19   &   1.189 &    1.371&  $ 5f6d7s $&$ ^2G $&$  ^2G  $& 7/2 &  8681 &     9202  &  0.911  &  0.899 &  0.889 \\
          &       &        &    &       &         &         &         &         &  $ 5f6d7s $&$ ^4F $&$  ^4F  $& 7/2 &  9304 &     9720  &  1.173  &  1.167 &  1.238 \\
$ 6d^27s $&$ ^4F $&$  ^4F $&7/2 &  4374 &   4147  &  1.232  &   1.227 &    1.238&  $ 5f6d7s $&$ ^4G $&$  ^4G  $& 7/2 & 10751 &    11117  &  0.983  &  0.977 &  0.984 \\
$ 6d^27s $&$ ^2G $&$  ^4G $&7/2 & 10502 &   9712  &  0.953  &   0.947 &    0.984&  $ 5f6d^2 $&$     $&$  ^2F  $& 7/2 & 13270 &    12486  &  0.855  &  1.036 &  1.143 \\
$ 6d^3   $&$     $&$  ^2F $&7/2 & 11621 &  10855  &  1.166  &   1.171 &    1.143&            &             &         &     &       &           &         &              \\
$ 6d^3   $&$     $&$  ^2F $&7/2 & 13297 &  12570  &  1.131  &   1.122 &    1.143&  $ 5f6d7s $&$     $&$  ^4H  $& 9/2 &  6265 &     6700  &  1.018  &  1.025 &  0.970 \\
$ 6d^3   $&$ ^2G $&$  ^2G $&7/2 & 18247 &  16818  &  0.916  &   0.906 &    0.889&  $ 5f6d7s $&$     $&$  ^2G  $& 9/2 &  8818 &     9238  &  1.086  &  1.068 &  1.111 \\
$ 6d^3   $&$ ^2F $&$  ^2F $&7/2 & 25165 &  22834  &  1.12   &   1.132 &    1.143&  $ 5f6d7s $&$     $&$  ^4H  $& 9/2 & 10435 &    10572  &  0.931  &  0.927 &  0.970 \\
          &       &        &    &       &         &         &         &         &  $ 5f6d7s $&$     $&$  ^4F  $& 9/2 & 12076 &    12488  &  1.245  &  1.253 &  1.333 \\
$ 6d^27s $&$ ^4F $&$  ^4F $&9/2 &  6528 &   6213  &  1.312  &   1.309 &    1.333&  $ 5f6d7s $&$ ^4G $&$  ^4G  $& 9/2 & 13156 &    13469  &  1.185  &  1.178 &  1.172 \\
$ 6d^27s $&$     $&$  ^2G $&9/2 & 11158 &  10379  &  1.153  &   1.145 &    1.111&  $ 5f6d^2 $&$ ^4H $&$  ^2H  $& 9/2 & 14421 &    15243  &  1.00   &  0.839 &  0.909 \\
$ 6d^3   $&$ ^4F $&$  ^4F $&9/2 & 14217 &  13249  &  1.242  &   1.256 &    1.333&            &             &               &  &       &           &         &          \\
$ 6d^3   $&$ ^2H $&$  ^4H $&9/2 & 16585 &  15305  &  1.006  &   0.983 &    0.970&  $ 5f6d7s $&$ ^4H$&$  ^4H $&11/2  &  9953 &    10189  &  1.128  &  1.121 &  1.133 \\
$ 6d^3   $&$ ^2G $&$  ^2G $&9/2 & 21328 &  19880  &  1.08   &   1.075 &    1.111&  $ 5f6d7s $&$ ^4G$&$  ^2H $&11/2  & 14390 &    15350  &  1.267  &  1.085 &  1.091 \\
$ 5f^27s $&$     $&$  ^2H $&9/2 & 30481 &  25246  &  0.96   &   0.939 &    0.909&  $ 5f6d^2 $&$ ^4I$&$  ^4G $&11/2  & 15111 &    16565  &  0.98   &  1.261 &  1.273 \\
          &       &        &    &       &         &         &         &         &  $  5f6d^2$&$ ^4H$&$  ^4I $&11/2  & 16717 &    17771  &  1.10   &  0.966 &  0.965 \\
$ 6d^3   $&$ ^2H$&$  ^4H $&11/2& 19039 &  17727  &  1.09   &   1.086 &    1.091&             &      &        &       &         &        &         &      &        \\
$ 5f^27s $&$ ^4H$&$  ^2H $&11/2& 33216 &  27937  &  1.12   &   1.118 &    1.133&             &      &        &       &         &        &         &      &        \\
$ 5f^27s $&$ ^2H$&$  ^4H $&11/2& 35702 &  30485  &  1.08   &   1.085 &    1.091&             &      &        &       &         &        &         &      &        \\
$ 5f^27s $&$ ^4K$&$  ^2I $&11/2& 37981 &  32621  &  0.826  &   0.885 &    0.923&             &      &        &       &         &        &         &      &        \\
          &      &        &    &       &         &         &         &         &             &      &        &       &         &        &         &      &        \\
$ 5f^27s $&$ ^4H$&$  ^4H $&13/2& 35827 &  30549  &  1.23   &   1.221 &    1.231&             &      &        &       &         &        &         &      &        \\
$ 5f^26d $&$ ^4K$&$  ^4K $&13/2& 41108 &  35401  &  0.98   &   0.973 &    0.964&             &      &        &       &         &        &         &      &        \\
$ 5f^27s $&$ ^2I$&$  ^4I $&13/2& 42291 &  37575  &  1.088  &   1.120 &    1.108&             &      &        &       &         &        &         &      &        \\
\end{tabular}
\end{ruledtabular}
\end{table*}

\section{Excitation energies of Th, Th$^{+}$, and Th$^{2+}$}
Excitation energies of the 73 lowest states of neutral Th are listed in Table~\ref{th1}.
All energies are given relative to the $6d^27s^2$~$^3F_2$ ground state. Theoretical results calculated with the
CI+all-order method are
listed in columns labelled ``Present''.  The results are compared with
experimental energies and $g$-factors given in
Ref.~\cite{Thorium}. Some of the energy levels listed in
\cite{Thorium} are only identified by the total angular momentum $J$, and not by a
complete $LSJ$ term designation. Such designations are always approximate and
 sometimes ambiguous, as in cases of strong configuration mixing.
 \subsection{Term identification using $g$-factors}
 \label{term}
  All of the states
reported in Table~\ref{th1} have some admixture of configurations, leading to ambiguities in term identification.
 Thus, we list two sets of term designations in Table~\ref{th1}:
the term listed in \cite{Thorium}, and our term identification, which is based on comparing
the experimental $g$-factor with the Land\`e formula value given by Eq.~(\ref{eq-gnr}) as described below.
   The corresponding columns are
labelled ``\cite{Thorium}'' and ``Present''. First, we group all levels by $J$ and by parity.
 This leads to a relatively small number of possible terms for each level since there are
 only three configuration present for even levels and four configurations for
  odd levels for the energy range in Table~\ref{th1}. We then identify the term appropriate
  for each level using the agreement of the experimental $g$-factor with the Land\`e formula.
When several configurations have the same $J$ and parity, we also verify which configuration has the largest mixing coefficient. When no entry appears in a term column that
indicates that no term has been proposed. The $g_\textrm{nr}$ calculated using  Eq.~(\ref{eq-gnr})
  are given in columns ``nr''. The ``Present'' column reports the actual $g$-factor that we calculate for the state as a whole. In general, the calculated $g$-factors are in good agreement with the Land\`e formula, but there are exceptions in cases of strong configuration mixing.

Two interesting entries in Table~\ref{th1} are the levels $6d^3 7s$~$^5F_1$ and $6d^27s7p$~$^5F_1$,
 for which the column ``nr'' reports a
value of zero for the theoretical $g$-factor. That factor is the
Land\`e g-factor given by Eq.~(\ref{eq-gnr}), which (but for a small
correction due to the anomalous magnetic moment of the electron)
describes the gyromagnetic ratio of a single-configuration quantum state
with well-defined quantum numbers $L$, $S$ and $J$ (respectively orbital, spin and total electronic angular momentum, in units of the reduced Planck constant, $\hbar$).
A vanishing (or very small) g-factor for a level implies that its energy
is relatively insensitive to the presence of a magnetic field.
Such an attribute
is of interest in applications to atomic frequency standards and precision measurement,
so we comment briefly on the use of Eq.~(\ref{eq-gnr}) as a screen in the search for small $g$-factors.

From Eq.~(\ref{eq-gnr}) we find that $g = 0$ when
\begin{equation} \label{gzero}
3J\left(J+1\right) - L\left(L+1\right) + S\left(S+1\right) = 0,
\end{equation}

\noindent where $L \geq 0$ is an integer; $J$ and $S$ are both non-negative integers or half-integers;
and $\left|L-S\right| \leq J \leq L+S$, the ``triangularity condition'', is satisfied.
Solution of the constrained equation (\ref{gzero}) is an exercise in integer programming,
for which even linear examples fall in the NP-hard class of computational complexity \cite{Karp1972}.
However, since only relatively small values of $J$, $L$ and $S$ are relevant to the periodic table,
the practical solutions of Eq.~(\ref{gzero})can be found by searching tables of computed values of its left-hand side.

Most of these solutions are of the type valid for any $J$:
\begin{equation} \label{gzero_standard}
L = 2 J + 1; \hspace{6 pt}  S = J + 1,
\end{equation}
\noindent i.e. those with term designations $^{2J+3}(2J+1)_J$ for $J = 0, \frac{1}{2}, 1, \ldots$.  Some of these terms are listed in Table 3 and Table 4 of Ref.~\cite{moore71a}.

A number of such levels are identified in the NIST Atomic Spectra Database \cite{NIST-web}, where they are usually found to have g-factors of 0.01 or less. Examples include
\vspace{6 pt}

\begin{itemize}
  \item Nd I $4f^4 6s^2 \, \, ^5\mathrm{F}_1$
  \item Pm II  $4f^5 6s \, \, ^7\mathrm{H}_2^\circ$
  \item Pm II  $4f^5 6s \, \, ^5\mathrm{F}_1^\circ$
  \item Tb I $4f^8 5d 6s^2 \, \, ^6\mathrm{G}_{3/2}$
  \item Dy II $4f^{10} \left(^5\mathrm{I}\right)5d  \, \, ^6\mathrm{G}_{3/2}$
  \item Sm I $4f^6 5d \left(^8\mathrm{H}\right) 6s \, \, ^7\mathrm{H}_2$
  \item Sm I $4f^6 5d \, \left(^6\mathrm{H}\right) 6s \, \,^7\mathrm{H}_2$

\end{itemize}

\vspace{6 pt}

\noindent There are other solutions of Eq.~(\ref{gzero}) for values of $J$, $L$ and $S$ that do not satisfy
Eq.~(\ref{gzero_standard}), such as $J = 3/2$, $L = 11$, $S = 21/2$.  These all appear to be associated with highly excited states involving multiple open shells.

\subsection{Hyperfine Land\`e g-factors}

A similar method can be used to search for vanishing Land\`e g-factors $g_F$ for hyperfine levels. Ignoring contributions from the nuclear magneton and the electron anomaly, we have (see Ref.~\cite{budker}, p. 83)

\begin{equation}
g_F = g_J\frac{F(F+1)+ J(J+1) - I(I+1)}{2F(F+1)}
\end{equation}

\noindent where $g_J$ is the g-factor associated with the electrons, $J$ the net electronic angular momentum,  $F$ the total angular momentum and $I$ the angular momentum of the nucleus.  Thus $g_F = 0$ when
\begin{equation}
F(F+1)=I(I+1)-J(J+1)
\label{gfcond}
\end{equation}
\noindent subject to the triangularity condition between $I$, $J$, and $F$.

The simple expression of Eq. \ref{gfcond} belies the diversity of its solutions, as suggested in Table \ref{gFzeros}.  A number of the values of $I$ displayed there are associated with stable or long-lived nuclei.

\begin{table}
\caption{Simplest non-trivial solutions of Eq. (\ref{gfcond}) for  $F \leq 10$.  The only solutions for $F <3/2$ are trivial.  Multiple solutions are given for $F = 4,5$ and 8 because they lie close together \label{gFzeros} }
    \begin{tabular}{ | c | c | c |c| c | c | c|}
    \hline
    $F$ & $J$ & $I$  &    &$F$ & $J$ & $I$ \\ \hline
3/2 &	3	 & 7/2     &   &6	 &	5 & 8  \\
2 &	 2 & 3         &  & 13/2	 &	8 &	 21/2 \\
5/2 &	8	 &	17/2   &  &  7	 &	9/2	 &	17/2 \\
3	 &	5	 &	6       &   & 15/2 &		11 &		27/2 \\
7/2 &		4	 &	11/2 &     &8 &		5/2	 &	17/2 \\
4 &	7/2	 &	11/2       &&8 &		13/2 &	21/2 \\
4  &		9 &	10      &   & 8  &	10 &		13 \\
9/2	 &	7 	 &	17/2     &&17/2	 &	80	 &	171/2 \\
5  &		3 &		6     &     &9 	& 6	 &	11 \\
5 &	 14 &		15         &&19/2 &	12	 &	31/2 \\
11/2 &	35	 &	71/2     &&10	 &	8  &	13 \\ \hline
    \end{tabular}
\end{table}

\vspace{12 pt}

 \begin{table*}
\caption{\label{th3}
 Levels (cm$^{-1}$) and
$g$-factors of the lowest states of two-times ionized
  thorium. Non-relativistic values of $g$-factors ($g_{\rm nr}$)
  are given by Eq.(\ref{eq-gnr}).
    Comparison of calculations with experiment \cite{RedNavSan14}. Configuration and term labels are determined as described in Section~\ref{term}.}
\begin{ruledtabular}
\begin{tabular}{lllllrrrrlllllrrrrr}
\multicolumn{1}{l}{Conf.}&
\multicolumn{3}{c}{Term}&
\multicolumn{2}{c}{Energy}&
\multicolumn{3}{c}{g-factors}&
\multicolumn{1}{c}{Conf.}&
\multicolumn{3}{c}{Term}&
\multicolumn{2}{c}{Energy}&
\multicolumn{3}{c}{g-factors}\\
\multicolumn{1}{c}{}&
\multicolumn{1}{c}{\cite{Thorium}}&
\multicolumn{1}{c}{Present}&
\multicolumn{1}{c}{$J$}&
\multicolumn{1}{c}{Present}&
\multicolumn{1}{c}{Expt.}&
\multicolumn{1}{c}{Expt.}&
\multicolumn{1}{c}{Present}&
\multicolumn{1}{c}{nr}&
\multicolumn{1}{c}{}&
\multicolumn{1}{c}{\cite{Thorium}}&
\multicolumn{1}{c}{Present}&
\multicolumn{1}{c}{$J$}&
\multicolumn{1}{c}{Present}&
\multicolumn{1}{c}{Expt.}&
\multicolumn{1}{c}{Expt.}&
\multicolumn{1}{c}{Present}&
\multicolumn{1}{c}{nr}\\
\hline
\multicolumn{9}{c}{Odd-parity states ($5f6d+5f7d+5f7s+5f8s+6d6f+6d7p $)}&
\multicolumn{9}{c}{Even-parity states ($6d^{2}+5f^{2}+7s^{2}+5f7p+5f6f+6d7s$)}\\
\hline
  $  5f6d $&$   ^3H$&$  ^3H$& 4&     0&      0.0 &  0.885   & 0.894 &    0.800  &$ 6d^2  $&$   ^3P$&$     $& 0&  6151&    5090  &  0.00    & 0.000 &    0.000\\
  $  5f6d $&$   ^3F$&$  ^1G$& 4&  3207&     3188 &  0.976   & 0.949 &    1.000  &$ 7s^2  $&$   ^1S$&$     $& 0& 12428&    11961 &  0.00    & 0.000 &    0.000\\
  $  5f7s $&$   ^3F$&$  ^3F$& 4&  6237&     6311 &  1.22    & 1.205 &    1.250  &$ 6d^2  $&$   ^1S$&$     $& 0& 22008&    18993 &  0.00    & 0.000 &    0.000\\
  $  5f6d $&$   ^3G$&$  ^3G$& 4&  8197&     8142 &  1.103   & 1.096 &    1.050  &$ 5f^2  $&$   ^3P$&$     $& 0& 29579&    29299 &  0.00    & 0.000 &    0.000\\
  $  5f6d $&$   ^3H$&$  ^3F$& 4&  9063&     8981 &  1.188   & 1.178 &    1.250  &$ 5f^2  $&$   ^1S$&$     $& 0& 55356&    51161 &  0.00    & 0.000 &    0.000\\
  $  6d7p $&$      $&$  ^3F$& 4& 54333&    53052 &  1.27    & 1.243 &    1.250  &$       $&$      $&$     $& 0& 84509&          &  0.00    & 0.000 &    0.000\\
  $  5f7d $&$      $&$  ^3H$& 4& 79062&    78417 &          & 0.891 &    0.800  &$ 5f6f  $&$      $&$     $& 0& 89234&    88313 &  0.00    & 0.000 &    0.000\\
  $  5f8s $&$      $&$  ^3F$& 4& 79519&    78930 &          & 1.210 &    1.250  &$       $&$      $&$     $& 0& 94112&          &  0.00    & 0.000 &    0.000\\
           &        &       &  &      &          &          &       &           &         &        &       &  &      &          &          &       &         \\
  $  5f6d $&$  ^3P$&$      $&0 & 11766&    11233 &    0.0   & 0.000 &    0.000  &$  6d7s $&$ ^3D  $&$  ^3D$&1 &  6137&     5524 &  0.50    & 0.499 &    0.500\\
  $  7s7p $&$  ^3P$&$      $&0 & 43188&    42260 &    0.0   & 0.000 &    0.000  &$  6d^2 $&$ ^3P  $&$  ^3P$&1 &  8905&     7876 &  1.50    & 1.491 &    1.500\\
  $  6d7p $&$     $&$      $&0 & 52776&    51745 &    0.0   & 0.000 &    0.000  &$  5f^2 $&$ ^3P  $&$  ^3P$&1 & 30636&    30402 &  1.494   & 1.492 &    1.500\\
  $  5f7d $&$     $&$      $&0 & 81644&    80906 &    0.0   & 0.000 &    0.000  &$  5f7p $&$      $&$  ^3D$&1 & 44946&    44603 &  0.495   & 0.497 &    0.500\\
           &       &        &  &      &          &          &       &           &         &        &       &  &      &          &          &       &         \\
  $  5f6d $&$   ^3D$&$  ^3D$& 1&  8260&     7921 & 0.621    & 0.599 &    0.500  &$  6d^2 $&$  ^3F $&$  ^3F$&2 &   895&       63 &  0.744   & 0.756 &    0.667\\
  $  5f6d $&$   ^3P$&$  ^3P$& 1& 11564&    11123 & 1.352    & 1.376 &    1.500  &$  6d^2 $&$      $&$  ^1D$&2 &  5426&     4676 &  1.020   & 1.003 &    1.000\\
  $  5f6d $&$   ^1P$&$  ^1P$& 1& 22733&    20711 &          & 1.006 &    1.000  &$  6d7s $&$  ^3D $&$  ^3D$&2 &  7943&     7176 &  1.180   & 1.155 &    1.167\\
  $  6d7p $&$      $&$  ^1P$& 1& 40282&    39281 & 0.911    & 0.902 &    1.000  &$  6d^2 $&$  ^3P $&$  ^3P$&2 & 11417&    10440 &  1.36    & 1.370 &    1.500\\
  $  7s7p $&$   ^3P$&$  ^3P$& 1& 45938&    45064 & 1.120    & 1.109 &    1.500  &$  6d7s $&$  ^1D $&$  ^1D$&2 & 16438&    16037 &  1.00    & 0.961 &    1.000\\
  $  6d7p $&$      $&$  ^3P$& 1& 51875&    50993 & 1.22     & 1.202 &    1.500  &$  5f^2 $&$  ^3F $&$  ^3F$&2 & 18616&    18864 &  0.694   & 0.738 &    0.667\\
  $  6d7p $&$      $&$  ^3P$& 1& 55035&    53939 & 1.25     & 1.248 &    1.500  &$  5f^2 $&$  ^1D $&$  ^3D$&2 & 28971&    28233 &  1.12    & 1.164 &    1.167\\
  $  7s7p $&$   ^1P$&$  ^1P$& 1& 69930&    69001 &          & 1.012 &    1.000  &$  5f^2 $&$  ^3P $&$  ^3P$&2 & 33488&    32867 &  1.344   & 1.300 &    1.500\\
           &        &       &  &      &          &          &       &           &         &        &       &  &      &          &          &       &         \\
  $  5f6d $&$   ^3F$&$  ^3F$& 2&   189&      511 & 0.711    & 0.765 &    0.667  &$  6d^2 $&$  ^3F $&$  ^3F$&3 &  4938&     4056 &  1.083   & 1.078 &    1.083\\
  $  5f6d $&$   ^3F$&$  ^3F$& 2&  2958&     3181 & 0.725    & 0.732 &    0.667  &$  6d7s $&$  ^3D $&$  ^3D$&3 & 10641&     9954 &  1.339   & 1.325 &    1.333\\
  $  5f6d $&$     J$&$  ^1D$& 2&  5797&     6288 & 0.908    & 0.858 &    1.000  &$  5f^2 $&$  ^3F $&$  ^3F$&3 & 20378&    20840 &  1.096   & 1.078 &    1.083\\
  $  5f6d $&$   ^3D$&$  ^3D$& 2& 10458&    10181 & 1.19     & 1.181 &    1.167  &$  5f7p $&$     $&$  ^3G$&3 & 33715&    33562 &  0.849   & 0.837 &    0.750\\
  $  5f6d $&$   ^3P$&$  ^3P$& 2& 13513&    13208 & 1.432    & 1.436 &    1.500  &$  5f7p $&$     $&$  ^3F$&3 & 38736&    38432 &  1.170   & 1.149 &    1.083\\
  $  6d7p $&$      $&$  ^3F$& 2& 38322&    37280 & 0.795    & 0.782 &    0.667  &$  5f7p $&$     $&$  ^1F$&3 & 42544&    42313 &  0.971   & 0.960 &    1.000\\
  $  6d7p $&$      $&$  ^3D$& 2& 45159&    44088 & 1.200    & 1.187 &    1.167  &$  5f7p $&$     $&$  ^3F$&3 & 47876&    47472 &  1.202   & 1.199 &    1.083\\
  $  6d7p $&$      $&$  ^1D$& 2& 48723&    47680 & 1.02     & 1.002 &    1.000  &$  6d7d $&$     $&$  ^1F$&3 & 84838&    83702 &          & 0.906 &    1.000\\
           &        &       &  &      &          &          &       &           &         &            &       &  &      &          &          &       &        \\
  $  5f7s $&$       $&$  ^3F$& 3&  2436&     2527 &  1.071  & 1.052 &    1.083  &$  6d^2 $&$  ^3F$&$  ^3F$&4 &  7264&     6538 &  1.20    & 1.161 &    1.250\\
  $  5f6d $&$       $&$  ^1F$& 3&  4853&     4827 &  1.003  & 0.988 &    1.000  &$  6d^2 $&$  ^1G$&$  ^3G$&4 & 10822&    10543 &  1.05    & 1.064 &    1.050\\
  $  5f6d $&$   ^3G $&$  ^3G$& 3&  5085&     5061 &  0.869  & 0.858 &    0.750  &$  5f^2 $&$  ^3H$&$  ^3H$&4 & 14514&    15149 &  0.81    & 0.814 &    0.800\\
  $  5f7s $&$       $&$  ^1F$& 3&  7609&     7501 &  1.027  & 1.015 &    1.000  &$  5f^2 $&$  ^3F$&$  ^3F$&4 & 21782&    21784 &  1.18    & 1.201 &    1.250\\
  $  5f6d $&$   ^3D $&$  ^3D$& 3& 11236&    10741 &  1.22   & 1.244 &    1.333  &$  5d2  $&$  ^1G$&$  ^1G$&4 & 27045&    25972 &  1.072   & 1.033 &    1.000\\
  $  5f6d $&$   ^1F $&$  ^3F$& 3& 16506&    15453 &  1.07   & 1.060 &    1.083  &$  5f7p $&$     $&$  ^3G$&4 & 38980&    38581 &  1.105   & 1.100 &    1.050\\
  $  6d7p $&$       $&$  ^3F$& 3& 45686&    44465 &  1.125  & 1.121 &    1.083  &$  5f7p $&$     $&$  ^3G$&4 & 44034&    43702 &  1.069   & 1.068 &    1.050\\
  $  6d7p $&$       $&$  ^3D$& 3& 51059&    49981 &  1.19   & 1.188 &    1.333  &$  5f7p $&$     $&$  ^3F$&4 & 47745&    47261 &  1.14    & 1.114 &    1.250\\
           &         &       &  &      &          &         &       &           &         &       &       &  &      &          &          &       &         \\
  $  5f6d $&$   ^3H$&$  ^3H$& 5&  4802&     4490 & 1.04     & 1.028 &    1.033  &$  5f^2 $&$ ^3H $&$  ^3H$&5 & 17131&    17888 &  1.01    & 1.028 &    1.033\\
  $  5f6d $&$   ^3G$&$  ^3G$& 5& 11456&    11277 & 1.186    & 1.187 &    1.200  &$  5f7p $&$     $&$  ^3G$&5 & 47781&    47422 &  1.207   & 1.194 &    1.200\\
  $  5f6d $&$   ^1H$&$  ^1H$& 5& 20144&    19009 & 1.001    & 1.001 &    1.000  &$  5f6f $&$     $&$  ^3I$&5 & 87443&    86934 &          & 0.887 &    0.833\\
  $  5f7d $&$      $&$  ^1H$& 5& 80841&    80137 &          & 1.033 &    1.000  &$  5f6f $&$     $&$  ^1H$&5 & 88176&    87667 &          & 1.006 &    1.000\\
  $  5f7d $&$      $&$  ^3G$& 5& 83847&    83023 &          & 1.137 &    1.200  &$  6d7d $&$     $&$  ^3G$&5 & 90979&    90085 &          & 1.141 &    1.200\\
  $  5f7d $&$      $&$  ^3H$& 5& 85158&    84239 &          & 1.047 &    1.033  &$  5f6f $&$     $&$  ^3H$&5 & 92840&    92103 &          & 1.020 &    1.033\\
  $  6d6f $&$      $&$  ^3H$& 5& 97802&    96317 &          & 1.049 &    1.033  &$  5f8p $&$     $&$  ^3G$&5 & 94046&    94144 &          & 1.186 &    1.200\\
           &        &       &  &      &          &          &       &           &         &        &      &  &      &          &          &       &         \\
  $  5f6d $&$   ^3H$&$  ^3H$& 6&  8810&     8437 &  1.17    & 1.160 &    1.167  &$  5f^2 $&$ ^3H $&$  ^3H$&6 & 20123&    20771 & 1.16     & 1.157 &    1.167\\
  $  5f7d $&$      $&$  ^3H$& 6& 84721&    83963 &          & 1.161 &    1.167  &$  5f^2 $&$ ^1I $&$  ^1I$&6 & 28635&    28350 &          & 0.999 &    1.000\\
  $       $&$      $&$  ^3H$& 6&102352&          &          & 1.160 &    1.167  &$  5f6f $&$     $&$  ^3I$&6 & 88845&    88387 &          & 1.021 &    1.024\\
  $       $&$      $&$  ^3I$& 6&108394&          &          & 1.032 &    1.024  &$  5f6f $&$     $&$  ^3H$&6 & 93648&    93045 &          & 1.101 &    1.167\\
  $       $&$      $&$  ^3K$& 6&108614&          &          & 0.868 &    0.857  &$  5f6f $&$     $&$  ^3I$&6 & 94584&    94018 &          & 1.052 &    1.024\\
  \end{tabular}
\end{ruledtabular}
\end{table*}

 \begin{table*}
\caption{\label{comp}
 Levels  (cm$^{-1}$)  of the lowest states of
 in one-time and two-times ionized  thorium. All energy values are given relative to the
 respective ground state.
    Comparison of calculations with experiment \cite{RedNavSan14} and other theory
    \cite{BerDzuFla09}.}
\begin{ruledtabular}
\begin{tabular}{llrrrllrrr}
\multicolumn{1}{c}{Th$^+$}&
\multicolumn{1}{c}{Term}&
\multicolumn{3}{c}{Energies}&
\multicolumn{1}{c}{Th$^{2+}$}&
\multicolumn{1}{c}{Term}&
\multicolumn{3}{c}{Energies}\\
\multicolumn{2}{c}{}&
\multicolumn{1}{c}{Present}&
\multicolumn{1}{c}{Expt. \cite{RedNavSan14}}&
\multicolumn{1}{c}{Th. \cite{BerDzuFla09}}&
\multicolumn{2}{c}{}&
\multicolumn{1}{c}{Present}&
\multicolumn{1}{c}{Expt. \cite{RedNavSan14}}&
\multicolumn{1}{c}{Th. \cite{BerDzuFla09}}\\
\hline
$5f7s^2 $&$  ^2F_{5/2}  $&  3882&   4490&    4856&  $6d^2  $&$ ^3F_{3} $&  4938 & 4056  &  4023 \\
$5f6d7s $&$  ^4F_{3/2}  $&  6020&   6691&    7487&  $6d^2  $&$ ^3F_{4} $&  7264 & 6538  &  6795 \\
$5f6d7s $&$  ^4F_{5/2}  $&  6651&   7331&    8325&  $6d7s  $&$ ^3D_{3} $& 10641 & 9954  &  9204 \\
$5f6d7s $&$  ^4G_{5/2}  $&  9229&   9585&   10045&  $6d^2  $&$ ^1G_{4} $& 10822 &10543  & 11051 \\
$5f6d7s $&$  ^4H_{5/2}  $&  9978&  10673&   12168&  $5f^2  $&$ ^3H_{4} $& 14514 &15149  & 13358 \\
$5f6d7s $&$  ^2D_{3/2}  $& 10695&  11576&   13054&  $5f^2  $&$ ^3H_{5} $& 17131 &17887  & 16068 \\
$5f6d7s $&$  ^4D_{1/2}  $& 11575&  11725&   12897&  $5f^2  $&$ ^3F_{3} $& 20378 &20840  & 19080 \\
$5f6d7s $&$  ^2F_{5/2}  $& 12045&  12472&   14564&  $5f^2  $&$ ^3F_{4} $& 21782 &21784  & 20366 \\
$5f6d7s $&$  ^4F_{3/2}  $& 12657&  12902&   14233&  $5f^2  $&$ ^1G_{4} $& 27045 &25972  & 25269 \\
$5f6d7s $&$  ^4G_{1/2}  $& 13965&  14102&   15853&  $5f7p  $&$ ^3G_{3} $& 33715 &33562  & 33402 \\
         &              &       &       &        &  $5f7p  $&$ ^3F_{3} $& 38736 &38432  & 38617 \\
  \end{tabular}
\end{ruledtabular}
\end{table*}

\subsection{Energies of neutral Th}
The calculated energy levels of thorium are in excellent agreement with experiment for such heavy tetravalent neutral atom.
We find
0.2 - 2\% differences between the theoretical and experimental
energies for 35 out of 78 levels listed in Table~\ref{th1}. Only 9 energies differ with experiment for more than
5\%.
 We find that inclusion of a sufficient number of configurations 
 is particularly important for achieving accurate results. To ensure that
 all dominant configurations are included we started with the preliminary calculation that included a
  few thousand configuration state functions (CSF). The initial configuration space was constructed by allowing two excitations into the valence orbitals from the $6d^27s^{2}$, $6d^37s$, $6d^27s7p$, $6d7s^{2}7p$, and $5f6d7s^{2}$ configurations. The results of that calculation allowed us to sort the configurations by their
  contributions to the energies of interest. Then, we allowed two more excitations into the restricted valence space from the $\sim$35 most
  important configurations and one more excitation into a large valence space from the $\sim$315 most important configurations.
  The restricted valence space includes  all orbitals with $l<5$ up to $10g$ and the large valence space includes orbitals up to $20d$.
  The combined file that includes all of the configurations of the initial run is constructed and duplicate entries are removed. The resulting list includes 24673 even and 28651 odd CSFs.
  This algorithm for construction of the configuration space was tested previously on tetravalent Hf~\cite{DzuSafSaf14}, where the results of this approach were compared with
  results  of much larger calculations.

  The differences between the preliminary and final runs are small only for a first few even states, ranging from 0 to 300~cm$^{-1}$. For most of the other states, the differences range between 900 and 1500~cm$^{-1}$, with the average difference being 1250~cm$^{-1}$. Moreover,
the results of our preliminary runs show that lack of saturation of the CI space results in 1500-2000~cm$^{-1}$ shift of all $5f6d7s^2$ and $5f6d^27s$ configurations relative to the even
$7s^26d^2$ configuration. Selective expansion of the configuration space fixes this problem resulting in the very good agrement between the energy levels of the  $5f6d7s^2$ and $5f6d^27s$ configurations with experiment.

The shift of the even vs. odd configurations involving an $nf$ state is a
well-known problem.
For example, the energies of the $4f^65d$
levels in Gd~IV  \cite{DzuSusJoh02}  were shifted by 13500~cm$^{-1}$ relative to the
$4f^7\ ^8S_{7/2}$ ground level to account for this problem. The procedure for correcting the shift
of the $4f^n5d$ energies relative to the ground state $4f^{n+1}$
energies was used for Nd~IV, Pm~IV, Sm~IV, and Eu~IV ions
\cite{DzuSafJoh03}.
\subsection{Energies of Th ions}

\begin{table*}
\caption{\label{tab-life-2014}  Lifetimes $\tau^{\rm  CI+all}$ (in seconds),
transition rates $A_r$ (in s$^{-1}$), and reduced
matrix elements $Z^{\rm CI+all}$ (in a.u.) for electric-multipole (E1 and
E2) and magnetic-multipole (M1 and M2) transitions in Ra-like
Th$^{2+}$ ion evaluated in the CI+all approximation.  Energies  (cm$^{-1}$) are from Ref.~\cite{Thorium}.
 The numbers in brackets represent powers of 10.}
\begin{ruledtabular}
\begin{tabular}{llrrrrrrrr}
\multicolumn{2}{c}{Transition}&
\multicolumn{1}{c}{}&
\multicolumn{2}{c}{Energies (cm$^{-1}$)}&
\multicolumn{1}{c}{$\lambda$}&
\multicolumn{1}{c}{$Z^{\rm no RPA}$}&
\multicolumn{1}{c}{$Z^{\rm CI+all}$}&
\multicolumn{1}{c}{$A_{r}^{\rm  CI+all}$}&
\multicolumn{1}{c}{$\tau^{\rm  CI+all}$}\\
\multicolumn{1}{c}{Upper}&
\multicolumn{1}{c}{Lower}&
\multicolumn{1}{c}{}&
\multicolumn{1}{c}{Lower}&
\multicolumn{1}{c}{Upper}&
\multicolumn{1}{c}{\AA}&
\multicolumn{1}{c}{a.u.}&
\multicolumn{1}{c}{a.u.}&
\multicolumn{1}{c}{s$^{-1}$}&
\multicolumn{1}{c}{sec}\\
\hline
 $ 6d^2 \ ^3F_{2}$&  $ 5f6d \  ^3H_{4} $&$M2$&     0.00 &    63.27 & 1580528&   2.1134 &   1.4359 &    6.224[-19]& 1.607[+18]\\[0.3pc]

 $ 5f6d \ ^3F_{2}$&  $ 5f6d \  ^3H_{4} $&$E2$&     0.00 &   510.76 &  195787&   3.6338 &   3.1838 &  7.901[-09]  & 1.266[+08]\\[0.3pc]

 $ 5f7s \ ^3F_{3}$&  $  6d^2\  ^3F_{2} $&$E1$&    63.27 &  2527.09 &   40587&   1.1061 &   0.7668 &  2.546[+03]  & 3.928[-04]\\[0.3pc]

 $ 5f7s \ ^3F_{2}$&  $ 5f6d \  ^3H_{4} $&$E2$&     0.00 &  3181.50 &   31432&  11.1410 &  10.6310 &  8.251[-04]  &3.038[+02]\\
                  &  $ 5f7s \  ^3F_{3} $&$M1$&  2527.09 &  3181.50 &  152809&   1.4806 &   1.4766 &  3.292[-03]  & \\[0.3pc]

 $ 5f6d \ ^1G_{4}$&  $ 5f6d \  ^3H_{4} $&$E2$&     0.00 &  3188.30 &   31365&   5.4527 &   5.0061 &  1.028[-04]  &1.740[+03]\\
                  &  $ 5f6d \  ^3F_{2} $&$E2$&   510.87 &  3188.30 &   37349&   3.5556 &   3.4795 &  2.073[-05]  &        \\
                  &  $ 5f7s \  ^3F_{3} $&$M1$&  2527.09 &  3188.50 &  151192&   0.7239 &   0.7220 &  4.513[-04]  & \\ [0.3pc]

 $  6d^2\ ^3F_{3}$&  $ 5f6d \  ^3H_{4} $&  $E1$&     0.00 &  4056.02 &   24655&   0.0662 &   0.0214 &  8.841[+00]  & 1.235[-03]\\
                  &  $ 5f6d \  ^3F_{2} $&  $E1$&   510.76 &  4056.02 &   28207&   0.3400 &   0.2483 &  7.953[+02]  &\\ [0.3pc]

$ 5f6d \ ^3H_{5}$&  $ 5f6d \  ^3H_{4} $&  $M1$&     0.00 &  4489.64 &   22274&   2.4360 &   2.4438 &  1.325[+00]  & 7.410[-01]\\
                 &  $ 5f6d \  ^1G_{4} $&  $M1$&  3188.30 &  4489.64 &   76844&   2.1292 &   2.1303 &  2.453[-02]  & \\ [0.3pc]

 $ 5f6d \ ^3F_{3}$&  $  6d^2\  ^3F_{2} $&   $E1$&    63.27 &  4826.83 &   20993&   1.1113 &   0.7438 &  1.731[+04]  &  5.777[-05]\\[0.3pc]
 $  6d^2\ ^3F_{2}$&  $ 5f7s \  ^3F_{3} $&   $E1$&  2327.09 &  4676.43 &   42565&   0.9413 &   0.5569 &  1.629[+03]  & 6.139[-04]\\[0.3pc]
 $ 5f6d \ ^3G_{3}$&  $  6d^2\  ^3F_{2} $&   $E1$&    63.27 &  5061.54 &   20007&   2.1735 &   1.4623 &  7.766[+04]  & 1.288[-05]\\[0.3pc]
 $  6d^2\ ^3P_{0}$&  $  6d^2\  ^3F_{2} $&   $E2$&    63.27 &  5090.06 &   19893&   5.8665 &   5.3985 &  1.047[-02]  & 9.555[+01]\\[0.3pc]

 $  6d7s\ ^3D_{1}$&  $ 5f6d \  ^3F_{2} $&   $E1$&   510.76 &  5523.88 &   19948&   0.7040 &   0.4896 &  2.040[+04]  &4.159[-05]\\
                  &  $ 5f7s \  ^3F_{2} $&   $E1$&  3181.50 &  5523.88 &   42692&   1.1268 &   0.6477 &  3.642[+03]  & \\[0.3pc]

 $ 5f6d \ ^1D_{2}$&  $  6d^2\  ^3F_{3} $&   $E1$&  4056.02 &  6288.42 &   44795&   0.1510 &   0.0727 &  2.379[+01]  & 4.187[-02]\\[0.3pc]
 $ 5f7s \ ^3F_{4}$&  $  6d^2\  ^3F_{3} $&   $E1$&  4056.02 &  6310.81 &   44350&   0.7621 &   0.5514 &  7.849[+02]  & 1.273[-03]\\[0.3pc]

 $  6d^2\ ^3F_{4}$&  $ 5f7s \  ^3F_{3} $&   $E1$&  2527.09 &  6537.78 &   24933&   0.2457 &   0.2126 &  6.568[+02]  & 1.131[-03]\\
                   &  $ 5f6d \  ^3F_{3} $&  $E1$&  4826.83 &  6537.81 &   58446&   0.3792 &   0.3273 &  1.208[+02]  &        \\
                   &  $ 5f6d \  ^3G_{3} $&  $E1$&  5060.94 &  6537.81 &   67711&   0.5535 &   0.3609 &  9.279[+01]  &  \\[0.3pc]

 $  6d7s\ ^3D_{2}$&  $ 5f7s \   ^3F_{3} $&  $E1$&  2527.09 &  7176.11 &   21510&   0.9399 &   0.5919 &  1.426[+04]  &6.720[-05]\\
                  & $ 5f6d \   ^3F_{3} $&  $E1$&  4826.83 &  7176.11 &   42566&   0.2150 &   0.3040 &   6.203[+02] & \\[0.3pc]

 $ 5f7s \ ^3F_{3}$&  $  6d^2\   ^3F_{2} $&   $E1$&    63.27 &   7500.61&   13446&   0.1114 &   0.0762 &  6.907[+02]  & 3.182[-04]\\
                 &  $  6d^2\   ^3F_{2} $&    $E1$&  4676.43 &  7500.61 &   35409&   0.8405 &   0.6070 &  2.403[+03]  &  \\
                 & $  6d^2\   ^3F_{4} $ &    $E1$&  6537.81 &  7500.61 &  103864&   0.4824 &   0.4054 &  4.249[+01]  &\\[0.3pc]

 $  6d^2\ ^3P_{1}$&   $ 5f6d \  ^3F_{2} $&  $E1$&   510.76 &  7875.83 &   13578&   0.1028 &   0.0840 &  1.902[+03]  & 4.898[-04]\\
                  &  $ 5f7s \  ^3F_{2} $&  $E1$&  3181.50 &  7875.83 &   21302&   0.0712 &   0.0446 &  1.392[+02]  & \\[0.3pc]

 $ 5f6d \ ^3G_{4}$&  $  6d^2\  ^3F_{3} $&  $E1$&  4056.02 &  8141.75 &   24475&   2.7360 &   1.8865 &  5.465[+04]  & 1.830[-05]\\[0.3pc]

 $  6d7s\ ^3D_{3}$&  $ 5f6d \   ^3H_{4} $&  $E1$&     0.00 &  9953.58 &   10047&   0.2032 &   0.1630 &  7.579[+03]  &2.666[-05]\\
                 &  $ 5f6d \   ^1G_{4} $&   $E1$&  3188.50 &  9953.58 &   14782&   0.4667 &   0.3538 &  1.122[+04]  &  \\
                 & $ 5f7s \   ^3F_{4} $&    $E1$&  6310.81 &  9953.58 &   27452&   1.6388 &   1.0983 &  1.687[+04]  &    \\[0.3pc]

  $  6d^2\ ^3P_{2}$&  $ 5f6d \   ^3F_{3} $& $E1$&  4826.83 & 10440.24 &   17814&   0.1880 &   0.1106 &  8.772[+02]  &1.982[-04]\\
                 &  $ 5f6d \   ^3G_{3}   $& $E1$&  5060.54 & 10440.24 &   18588&   0.2160 &   0.1468 &  1.353[+03]  &   \\           
                 & $ 5f7s \  ^3F_{3}     $&  $E1$&  7500.61 & 10440.24 &   34018&   0.7666 &   0.5082 &  2.658[+03]  &  \\[0.3pc]

 $  6d^2\ ^1G_{4}$ &  $ 5f6d \  ^3F_{3} $&    $E1$&  4826.83 & 10542.90 &   17495&   1.0369 &   0.7953 &  2.949[+04]  &2.366[-05] \\
                   & $ 5f6d \  ^3G_{3} $&     $E1$&  5060.54 & 10542.90 &   18240&   0.7825 &   0.5208 &  1.001[+04]  &           \\

 $  7s^2\ ^1S_{0}$&  $  6d^2\  ^3F_{2} $&   $E2$&  4676.43 & 11961.13 &   13727&   7.0555 &   6.7483 &  1.046[-01]  &  1.418[+00]\\
                  &  $  6d^2\  ^3P_{1} $&   $M1$&  7875.83 & 11961.13 &   24478&   0.5484 &   0.5504 &  5.572[-01]  &                 \\
\end{tabular}
\end{ruledtabular}
\end{table*}

\begin{table}
\caption{\label{tab-tran-e1}  Oscillator strengths $f$ and
transition rates $A_r$ (s$^{-1}$) for electric-dipole
 transitions in Ra-like
Th$^{2+}$ ion. Wavelengths
(\AA)  from  compilation of Ref.~\cite{Thorium} are
listed for reference.
 The numbers in brackets represent powers of 10.}
\begin{ruledtabular}
\begin{tabular}{llrrr}
\multicolumn{2}{c}{Transition}&
\multicolumn{1}{c}{Wavelength}&
\multicolumn{1}{c}{$f^{\rm  CI+all}$}&
\multicolumn{1}{c}{$A_{r}^{\rm  CI+all}$}\\
\multicolumn{1}{c}{Lower}&
\multicolumn{1}{c}{Upper}&
\multicolumn{1}{c}{\AA}& \multicolumn{1}{c}{}&
\multicolumn{1}{c}{s$^{-1}$}\\
\hline
 $ 5f6d \  ^3H_{4} $& $   6d^2 \  ^3G_{4} $&    9485.6&    3.733[-2] &    3.417[+4]\\
 $ 6d^2 \  ^3F_{2} $& $   5f6d \  ^3D_{1} $&   12726.7&    1.651[-2] &    4.532[+4]\\
 $ 5f7s \  ^3F_{3} $& $   6d7s \  ^3D_{3} $&   13465.8&    1.664[-2] &    1.250[+4]\\
  $ 5f6d \  ^3F_{2} $& $   6d7s \  ^3D_{2} $&   15003.5&    9.591[-3] &    1.137[+4]\\
 $ 5f6d \  ^3H_{4} $& $   6d^2 \  ^3F_{4} $&   15296.1&    6.768[-3] &    2.383[+3]\\
 $ 6d^2 \  ^1D_{2} $& $   5f6d \  ^3P_{1} $&   15512.2&    6.619[-3] &    1.223[+4]\\
 $ 6d^2 \  ^3F_{2} $& $   5f6d \  ^1D_{2} $&   16064.4&    1.080[-3] &    1.117[+3]\\
 $ 6d^2 \  ^3F_{3} $& $   5f6d \  ^3D_{2} $&   16327.7&    3.118[-2] &    2.229[+4]\\
 $ 6d^2 \  ^1D_{2} $& $   5f6d \  ^3D_{3} $&   16489.3&    6.448[-2] &    4.520[+4]\\
  $ 6d7s \  ^3D_{1} $& $   5f6d \  ^3P_{0} $&   17515.8&    4.814[-3] &    3.488[+4]\\
 $ 6d7s \  ^3D_{1} $& $   5f6d \  ^3P_{1} $&   17859.9&    4.407[-3] &    1.024[+4]\\
 $ 6d^2 \  ^3F_{2} $& $   5f6d \  ^3G_{3} $&   20011.4&    1.625[-1] &    7.737[+4]\\
 $ 6d^2 \  ^3F_{3} $& $   5f6d \  ^3F_{4} $&   20307.0&    9.844[-2] &    2.528[+4]\\
 $ 6d^2 \  ^3F_{4} $& $   5f6d \  ^3G_{5} $&   21102.0&    5.705[-1] &    8.633[+4]\\
  $ 6d^2 \  ^3F_{4} $& $   5f6d \  ^3D_{3} $&   23791.1&    1.139[-1] &    2.131[+4]\\
 $ 5f6d \  ^3F_{2} $& $   6d^2 \  ^1D_{2} $&   24006.2&    8.657[-3] &    4.007[+3]\\
  $ 5f6d \  ^3D_{1} $& $   7s^2 \  ^1S_{0} $&   24752.7&    7.869[-4] &    2.857[+3]\\
 $ 5f6d \  ^3F_{2} $& $   6d^2 \  ^3F_{3} $&   28207.2&    3.317[-3] &    7.947[+2]\\
 $ 5f6d \  ^1G_{4} $& $   6d^2 \  ^3F_{4} $&   29855.6&    2.675[-2] &    2.473[+3]\\
 $ 6d^2 \  ^3F_{2} $& $   5f6d \  ^3F_{2} $&   32070.0&    1.603[-3] &    4.159[+2]\\
  $ 6d^2 \  ^1D_{2} $& $   5f7s \  ^1F_{3} $&   35396.5&    1.569[-2] &    2.386[+3]\\
  $ 5f7s \  ^3F_{3} $& $   6d^2 \  ^1D_{2} $&   46526.4&    1.439[-2] &    1.266[+3]\\
 $ 6d^2 \  ^1D_{2} $& $   5f6d \  ^1D_{2} $&   62035.6&    1.691[-3] &    1.172[+2]\\
 $ 5f7s \  ^3F_{3} $& $   6d^2 \  ^3F_{3} $&   65405.7&    1.286[-2] &    4.094[+2]\\
 $ 6d^2 \  ^3F_{2} $& $   5f6d \  ^3F_{2} $&  223469.2&    3.053[-3] &    1.637[+1]\\
\end{tabular}
\end{ruledtabular}
\end{table}

\begin{table*}
\caption{\label{tab-life-comp}  Lifetimes
(ns), transition rates $A_r$ (s$^{-1}$), and branching ratios  of
electric-dipole transitions in Ra-like Th$^{2+}$. Experimental lifetimes are
taken from Ref.~\cite{BiePalQui02}.
 Levels
(cm$^{-1}$) are from the experimental compilation of Ref.~\cite{Thorium}.
 The numbers in brackets represent powers of 10.}
\begin{ruledtabular}
\begin{tabular}{llrrrrrrrrr}
\multicolumn{2}{c}{Transition}&
\multicolumn{2}{c}{Levels (cm$^{-1}$)}&
\multicolumn{1}{c}{$\lambda$}&
\multicolumn{1}{c}{$Z^{\rm CI+all}$}&
\multicolumn{1}{c}{$A_{r}^{\rm  CI+all}$}&
\multicolumn{1}{c}{Branch.}&
\multicolumn{1}{c}{$\tau^{\rm  CI+all}$}&
\multicolumn{1}{c}{$\tau^{\rm  Expt}$}\\
\multicolumn{1}{c}{Upper}&
\multicolumn{1}{c}{Lower}&
\multicolumn{1}{c}{Upper}&
\multicolumn{1}{c}{Lower}&
\multicolumn{1}{c}{\AA}&
\multicolumn{1}{c}{a.u.}&
\multicolumn{1}{c}{s$^{-1}$}&
\multicolumn{1}{c}{ratio}&
\multicolumn{1}{c}{ns}&
\multicolumn{1}{c}{ns}\\
\hline
   $5f^2 \  ^3P_{2} $& $    5f6d \  ^3F_{2}$&     32867.27&        510.76&    3090.6&    0.8137&    9.09[+6]&       0.19& 21.2& 25.8$\pm$1.5\\
   $5f^2 \  ^3P_{2} $& $    5f6d \  ^1D_{2}$&     32867.27&       6288.42&    3762.4&    0.7188&    3.93[+6]&       0.08&&               \\
   $5f^2 \  ^3P_{2} $& $    5f6d \  ^3D_{3}$&     32867.27&      10741.15&    4519.5&    1.0101&    4.48[+6]&       0.10&&               \\
   $5f^2 \  ^3P_{2} $& $    5f6d \  ^3P_{1}$&     32867.27&      11123.18&    4599.0&    0.8850&    3.26[+6]&       0.07&&               \\
   $5f^2 \  ^3P_{2} $& $    5f6d \  ^3P_{2}$&     32867.27&      13208.21&    5086.7&    2.1928&    1.48[+7]&       0.31&&               \\
   $5f^2 \  ^3P_{2} $& $    5f6d \  ^3F_{3}$&     32867.27&      15453.41&    5742.6&    1.6403&    5.76[+6]&       0.12&&               \\[0.4pc]
   $5f7p \  ^3G_{4} $& $    5f6d \  ^1G_{4}$&     38580.60&       3188.30&    2825.5&    2.0580&    4.23[+7]&       0.10&  2.41& 2.7$\pm$0.2\\
   $5f7p \  ^3G_{4} $& $    5f6d \  ^3H_{5}$&     38580.60&       4489.64&    2933.3&    3.1200&    8.68[+7]&       0.21&&               \\
   $5f7p \  ^3G_{4} $& $    5f7s \  ^3F_{4}$&     38580.60&       6310.81&    3098.9&    3.8614&    1.13[+8]&       0.27&&               \\
   $5f7p \  ^3G_{4} $& $    5f7s \  ^1F_{3}$&     38580.60&       7500.61&    3217.5&    3.2880&    7.31[+7]&       0.18&&               \\
   $5f7p \  ^3G_{4} $& $    5f6d \  ^3F_{4}$&     38580.60&       8980.56&    3378.4&    2.5555&    3.81[+7]&       0.09&&               \\[0.4pc]
   $7s7p \  ^3P_{0} $& $    6d7s \  ^3D_{1}$&     42259.71&       5523.88&    2722.1&    1.2104&    1.47[+8]&       0.91&  6.19& 6.6$\pm$0.4\\
   $7s7p \  ^3P_{0} $& $    6d^2 \  ^3P_{1}$&     42259.71&       7875.83&    2908.3&    0.4150&    1.42[+7]&       0.09&&               \\[0.4pc]
   $7s7p \  ^3P_{1} $& $    6d^2 \  ^3F_{2}$&     45063.97&         63.27&    2222.2&    1.0592&    6.91[+7]&       0.15&  2.22& 2.4$\pm$0.2\\
   $7s7p \  ^3P_{1} $& $    6d^2 \  ^1D_{2}$&     45063.97&       4676.43&    2476.0&    2.2857&    2.33[+8]&       0.52&&               \\
   $7s7p \  ^3P_{1} $& $    6d7s \  ^3D_{1}$&     45063.97&       5523.88&    2529.1&    0.7118&    2.12[+7]&       0.05&&               \\
   $7s7p \  ^3P_{1} $& $    6d7s \  ^3D_{2}$&     45063.97&       7176.11&    2639.4&    1.4716&    7.96[+7]&       0.18&&               \\
   $7s7p \  ^3P_{1} $& $    7s^2 \  ^1S_{0}$&     45063.97&      11961.13&    3020.9&    0.8471&    1.76[+7]&       0.04&&               \\ [0.4pc]
   $6d7p \  ^3F_{4} $& $    6d^2 \  ^3F_{3}$&     53052.47&       4056.02&    2041.0&    0.7944&    1.67[+7]&       0.02&  1.41 &1.3$\pm$0.2\\
   $6d7p \  ^3F_{4} $& $    6d^2 \  ^3F_{4}$&     53052.47&       6537.81&    2149.9&    2.3177&    1.22[+8]&       0.17&&               \\
   $6d7p \  ^3F_{4} $& $    6d7s \  ^3D_{3}$&     53052.47&       9953.58&    2320.2&    5.3100&    5.08[+8]&       0.72&&               \\
   $6d7p \  ^3F_{4} $& $    6d^2 \  ^3G_{4}$&     53052.47&      10542.90&    2352.4&    1.7656&    5.39[+7]&       0.08&&               \\[0.4pc]
\end{tabular}
\end{ruledtabular}
\end{table*}

The energies and $g$-factors of  38 even-parity and 31 odd-parity
states of trivalent Th$^+$ are listed in Table~\ref{th2}. The table is structured in the same way
as Table~\ref{th1}.
All values are counted from the  $6d^27s\ ^2D_{3/2}$ ground state
energy. We note that Ref.~\cite{Thorium} does not aasign the ground state $LSJ$ term designation.  The corresponding experimental and our calculated ground state $g$-factors
listed in the first line of Table~\ref{th2}, $g_{\rm Expt}$ = 0.639 and $g_{\rm Present}$ =
0.662, are in between  the nonrelativistic values 0.4 and 0.8 for the
 $^4F_{3/2}$ and  $^2D_{3/2}$ terms, respectively.
We assign the $^2D_{3/2}$ term designation to the ground state of Th$^+$.

The 38 even-parity states listed in Table~\ref{th2} belong to
 five configurations, $6d7s^2$, $6d^27s$, $6d^{3}$, $5f^27s$, and
$5f^26d$, which are strongly mixed. Most odd-parity levels shown in
Table~\ref{th2} have a predominant composition of $5f6d7s$.
 The CI+all-order results are
in good agreement with experiment for most states, with the
differences being  less than  5\% for
the $5f6d7s$ states.
The  larger discrepancies with experiment that are observed for even-parity levels with two $5f$ electrons, such as
  $5f^27s$ and $5f^26d$  are most likely due to effects of correlations involving  higher ($l>6$) partial waves. This problem is exacerbated when two $5f$ electrons are present in the same configuration.

Energies and $g$-factors for 95 levels of divalent
Th$^{2+}$ are listed in Table~\ref{th3}. The energies are counted from the ground state.
The energies of the
odd-parity $5f6d$, $5f7d$ $5f7s$, $5f8s$, $6d6f$, and  $6d7p$ configurations and
even-parity $6d^{2}$, $5f^{2}$, $7s^{2}$, $5f7p$, $5f6f$, and $6d7s$ configurations
calculated with the CI + all-order approach are compared with
experimental energies \cite{Thorium} given in columns ``Present''
and ``Expt'' of Table~\ref{th3}. We note that
 the ground state of Th$^{2+}$ is $5f6d\ ^3H_{4}$
instead of the usual $ns^2\ ^1S_0$, such as in isoelectronic Ac$^+$ and Ra.
The theoretical values agree well with experiment for most cases,
 with the exception of the $5f6d\
^3F_2$ and
$6d^2\ ^3F_2$ levels that are very close to the ground state. Since we calculate these
energies as the differences of the large ground and excited divalent removal energies, the accuracy is reduced for such small energy intervals.
Our values agree with experiment to 0.10\% - 1\% for 38 levels.

In Table~\ref{comp}, we compare all results for levels for which theoretical calculations were performed both by us and by Berengut et al.~\cite{BerDzuFla09}, where a CI+MBPT approach was used. Berengut et al.~\cite{BerDzuFla09} noted that
``while we believe the $6d^2\ ^3F_3$, $6d^2\ ^3F_4$, and
$6d7s\ ^3D_3$ transitions are accurate, the others are estimates
only.'' Comparison of the two theoretical results with experimental
data \cite{Thorium} given in Table~\ref{comp} shows that our
results for Th$^{2+}$  are in better agreement with
experiment  than the ``CI + MBPT''
values except for the $6d^2\ ^3F_3$ and
$6d^2\ ^3F_4$ levels.  The case of Th$^+$ is similar.
The CI+all-order method includes higher-order correlation beyond the
CI+MBPT approach. However, in some cases the higher-order terms may cancel with other contributions.

The CI+MBPT approach was used by Porsev and Flambaum
\cite{PorFla10} to evaluate energies and $g$-factors in Th$^{+}$.
Tabulated results were given for even-parity states with $J$ =3/2
and 5/2 in the range from 18119~cm$^{-1}$ to 40644~cm$^{-1}$. Since our
calculation was carried out for lower levels, we can only compare results for two states,
$6d^3\ ^2D_{3/2}$  and $6d^3\
^4D_{5/2}$. The CI+MBPT results of \cite{PorFla10} differ from the experimental values by 18\%
while our CI+all-order values by 10\%.

\section{Multipole transition amplitudes and lifetimes  in Ra-like Th$^{2+}$ }

We now discuss some multipole transition amplitudes in Th$^{2+}$ which are
representative of the calculations that may have to be done for the
investigation of the electronic bridge process \cite{PorFlaPei10}.
Our CI + all-order results for the  multipole matrix elements,
transition rates, and lifetimes in Ra-like Th$^{2+}$ are given
in Table~\ref{tab-life-2014}.

We evaluate multipole matrix elements between 12 odd-parity states with energies in the
0 - 8142~cm$^{-1}$ range with 12 even-parity states with energies in the
63 - 11961~cm$^{-1}$ range. This results in 45 E1, 66 M2, and 82 E3 transitions
between the odd-even and even-odd states. We evaluate also the 83
M1, E2, and M3 transitions inside of even-parity complex, as well
as the 110  M1, E2, and M3 transitions inside of odd-parity
complex. That gives us 386 multipole matrix elements for
transitions between lowest-lying levels in Th$^{2+}$ ion.

In Table ~\ref{tab-life-2014}, we include
results for 45 selected electric-multipole (E1 and E2) and magnetic-multipole
(M1 and M2) transitions that are most important for the evaluation of the
corresponding lifetimes.
The octupole (E3 and M3) transitions  make
negligible contributions to the
lifetimes and are omitted.

We use atomic units (a.u.) to express all transition matrix
elements throughout this section: the numerical values of the elementary
 charge, $e$, the reduced Planck constant, $\hbar = h/2
\pi$, and the electron mass, $m_e$, are set equal to 1.
The atomic unit for electric-dipole matrix element is  $ea_0$, where
$a_0$ is the Bohr radius.

To show the importance of using effective transition operators (for example electric-dipole
$D^{\textrm{eff}})$,
 which include random-phase-approximation (RPA)  corrections, instead of the ``bare'' operators, we give the matrix elements
with and without the RPA correction in  columns labelled
$Z^{\textrm{CI+all-order}}$ and $Z^{\textrm{no RPA}}$, respectively.
We find that the RPA correction is significant (20-50\%) for most transitions, so
the $Z^{\textrm{CI+all-order}}$ final values are used in calculating transition rates and lifetimes.
The RPA correction is small for M1 matrix elements, $5f7s\ ^3F_{3} -5f7s\
 ^3F_{2}$ and $6d^2\  ^3P_{1} -  7s^2\ ^1S_{0}$.

The E1, E2, E3,  M2, M3, and M3 transition probabilities $A_r$
(s$^{-1}$) are obtained in terms of line strengths $S$~(a.u.) and
energies $\mathcal{E}$~(a.u.) as
\begin{eqnarray}
A(Ek) &=&\frac{C^{(k)}\left[\mathcal{E} \right] ^{2k+1}}{(2J+1)}S(Ek),\
\label{tr1} \\
C^{(1)} &=& 2.14200\times 10^{10}, \nonumber\\
\nonumber
C^{(2)} &=& 5.70322\times 10^{4},\\
C^{(3)} &=& 7.71311\times 10^{-2}, \nonumber\\
\ A(Mk) &=&\frac{D^{(k)}\left[ \mathcal{E}\right] ^{2k+1}}{(2J+1)}S(Mk),  \\
D^{(1)} &=& 2.85161\times 10^{5}, \nonumber\\
D^{(2)} &=& 7.59260\times 10^{-1},\nonumber \\
D^{(3)} &=& 1.02683\times 10^{-6}. \nonumber
\end{eqnarray}

The line strengths $S(E1)$, $S(E2)$, $S(E1)$, $S(E3)$, $S(M1)$,
$S(M2)$, and $S(M3)$ are obtained as squares of the corresponding
 E1, E2, E3, M2, M3, and M3  matrix elements listed in  column
 $Z^{\rm CI+all}$ of Table~\ref{tab-life-2014}.
Energies  are from the experimental compilation of Ref.~\cite{Thorium}. We list
the experimental energies for upper and lower states as well as the corresponding transition
wavelengths $\lambda$ in
Table ~\ref{tab-life-2014} for reference.
Our results for the transition rates are given in column $A_{r}^{\rm  CI+all}$
of Table~\ref{tab-life-2014}.

In order to determine the lifetimes listed in the last column of
Table~\ref{tab-life-2014}, we sum over all possible radiative transitions.
 The number of
contributing transitions increases significantly for higher levels. For
example, 18 transitions contribute to the lifetime of the relatively low-lying $5f7s\
^3F_4$ state, $E(5f7s\ ^3F_4)$ = 6310.81~cm$^{-1}$. However,
only one transition, $6d^2\ ^3F_{3} - 5f7s\ ^3F_{4}$,
 contributes significantly, and the total contribution of other 17 transitions to
  the $5f7s\ ^3F_{4}$
lifetime  is equal to 0.1\%.
The final values of $\tau^{\rm  CI+all}$ for 23 lowest-lying
levels are listed in the last column of Table~\ref{tab-life-2014}.
The term designation for those levels are in the first column of
Table~\ref{tab-life-2014}.
 In Table~\ref{tab-tran-e1}, we present results for other E1 transitions
 for low-lying levels with smaller transition rates.

Unfortunately, we did not find any theoretical or experimental
results to compare with our $A_{r}$ and $\tau$ values for the  low-lying states listed in
 Table~\ref{tab-life-2014}.
 Experimental
measurements of lifetimes for six higher levels of Th$^{2+}$ were performed  by
Bi\'emont {\it et al.\/} \cite{BiePalQui02}, and are summarized in Table~\ref{tab-life-comp}.
In order to calculate these lifetimes with our CI+all-order method,
 we carried out extensive additional calculations to obtain the wave functions and
 corresponding  electric-dipole matrix elements for
 112 levels with $0 \leq J \leq 6$.
  Using the calculated
CI+all-order electric-dipole matrix elements and experimental energies
\cite{Thorium}, we obtain values for 1152 transition rates $A_r$.
These are summed over all transitions to determine the lifetimes for
about 100 levels.

 In
Table~\ref{tab-life-comp}, we present   lifetimes, $\tau^{\rm
CI+all}$,  transition rates $A_r$,  and branching ratios of
electric-dipole transitions relevant to  comparison with the
lifetime measurements of \cite{BiePalQui02}. Energies in both tables  are from  the
compilation of Ref.~\cite{Thorium}. Only the dominant transitions are
listed, but all transitions are  included in the lifetime
calculation.

In the two last columns of Table~\ref{tab-life-comp}, we compare our
CI+all-order lifetimes with measurements \cite{BiePalQui02}.  We find
 excellent agreement between the CI+all-order lifetimes and
 experimental values for the $5f7p\ ^3G_{4}$, $7s7p\ ^3P_{0}$,
 $7s7p\ ^3P_{1}$, and $6d7p\ ^3F_{4}$ levels.
There is a 15\% difference for the lifetime of the $5f^2\
^3P_{2}$ level. Several channels contribute significantly to this lifetime.
The largest  branching
ratio for this level is 0.31 for the $5f6d\ ^3P_{2} - 5f^2 \
^3P_{2}$ transition. The branching ratios for the other 5 transitions
shown in Table~\ref{tab-life-comp} add 56\%.  An additional 13\%
comes from eight transitions that are not shown in
Table\ref{tab-life-comp}.

\section{Conclusion}
In summary,  a systematic study of the
   Th, Th$^{+}$, and Th$^{2+}$ energies was carried out using the CI+all-order approach.
Excitation energies are compared with  experimental \cite{Thorium}
and with available theoretical results \cite{BerDzuFla09}.
Good agreement with experiment was found even for neutral Th owing to
all-order treatment of the dominant correlation corrections and sufficient saturation of the
 configuration space.
The differences between the theoretical and experimental
energies of neutral thorium did not exceed 2\% for 35 out of 78 levels listed in Table~\ref{th1}, and  only 9 energies differed with experiment for more than
5\%. These result show the success of the algorithm that we have developed for
efficient selection of important configurations for tetravalent systems.

We explored the issue of accidentally ``vanishing'' $g$-factors, where the Land\`e formula gives
$g=0$, such as for $^5F_1$ terms. We identified a number of cases in which hyperfine $g$-factors may also vanish.

The recommended values are provided for multipole transition rates and lifetimes of low-lying
levels  in Ra-like Th$^{2+}$. We expect these values to be accurate to a few percent for strong transitions based on
our calculations in divalent alkaline-earth atoms~\cite{sr}.
To further verify the accuracy of our calculations, we compared our results with the only available
experimental lifetimes \cite{BiePalQui02} for higher
$5f7p\ ^3G_{4}$, $7s7p\ ^3P_{0}$,
 $7s7p\ ^3P_{1}$, and $6d7p\ ^3F_{4}$ levels of Th$^{2+}$ ion.
This works demonstrates the ability to perform accurate calculations for Th and its ions for Th nuclear clock development and other applications.

\section*{Acknowledgment}
We thank Stephen Jordan and Joseph Reader for helpful comments.
This research was performed under the sponsorship of the US
Department of Commerce, National Institute of Standards and
Technology, and was supported by the National Science Foundation
under Physics Frontiers Center Grant No. PHY-0822671.

\end{document}